\documentclass[12pt,preprint,linenumbers]{article}

\usepackage{times}
\usepackage{lineno}
\usepackage{color}

\usepackage[superscript]{cite}

\usepackage{graphicx}
\usepackage{epstopdf}
\graphicspath{ {./new_figs/} }
\usepackage{amssymb}
\DeclareGraphicsExtensions{.eps,.jpg,.png,.pdf}
\usepackage{caption}
\usepackage{setspace}
\renewcommand{\figurename}{\bf Fig.}

\title{Revealing the three-dimensional structure of liquids
using four-point correlation functions}

\author
{Zhen Zhang,$^{1}$ Walter Kob,$^{1,2\ast}$ \\
\\
\normalsize{$^{1}$Laboratoire Charles Coulomb, University of Montpellier and CNRS,}\\
\normalsize{F-34095 Montpellier, France}\\
\normalsize{$^{2}$Institut Universitaire de France (IUF)}\\
\\
\normalsize{$^\ast$Correspondence to:  walter.kob@umontpellier.fr}
}

\date{}

\begin{document} 




\maketitle

\begin{abstract}
Disordered systems like liquids, gels, glasses, or granular materials are
not only ubiquitous in daily life and in industrial applications but they
are also crucial for the mechanical stability of cells or the transport
of chemical and biological agents in living organisms.  Despite the
importance of these systems, their microscopic structure is understood
only on a rudimentary level, thus in stark contrast to the case of gases
and crystals. Since scattering experiments and analytical calculations
usually give only structural information that is spherically averaged,
the three dimensional (3D) structure of disordered systems is basically
unknown. Here we introduce a simple method that allows to probe the 3D
structure of such systems. Using computer simulations we find that 
hard-sphere-like liquids have on intermediate and large scales an intricate
structural order given by alternating layers with icosahedral and
dodecahedral symmetries, while open network liquids like silica have a
structural order with tetrahedral symmetry. These results show that
liquids have a highly non-trivial 3D structure and that this structural
information is encoded in non-standard correlation functions.

\end{abstract}

\date{\today}

\newpage


The microscopic structure of many-particle systems is usually determined from
scattering experiments which give access to the static structure factor
$S(\vec{q})$, $\vec{q}$ is the wave-vector, and for crystals such
measurements allow to obtain a complete know\-ledge of the structure
of the material~\cite{hansen_86,salmon_06,ashcroft_76,binder_11}. This is not
the case for disordered materials such as liquids, foams, granular
materials, since on the macroscopic scale these are isotropic
and hence $S(\vec{q})$ depends only on the norm $q=|\vec{q}|$, i.e.,
the whole three dimensional structural information is projected onto
a single function $S(q)$. This projection entails a
huge loss of structural information, which subsequently
has to be recovered, at least partially, from physical
arguments on the possible arrangement of the particles. Such
arguments exist for the first few nearest neighbor shells of the
particles~\cite{jonsson_88,ma_11,miracle_04,wochner_09,xia_17,royall_15,malins_13,dunleavy_15,coslovich_07,tanaka_12,fang_10,fang_11},
but not for the arrangements on larger scales.

Microscopy on colloidal systems and computer simulations have shown that
for hard-sphere-like systems the local structure can be surprisingly
varied, in particular if the liquid is constituted by more than one type
of particle. The geometry of these locally favored structures depends on
packing fraction and is rather sensitive to parameters like the composition of
the system, the size of the particles, or the interaction energies~\cite{royall_15,ma_11}. As a
consequence it has so far not been possible to come up with a universal
description of the structure on the local scale and it is unlikely that
such a universal description exists.

In view of this difficulty it is not surprising that very little
effort has been made so far to investigate the structure of disordered
systems on length scales beyond the first few nearest neighbors~\cite{royall_15,fang_10}. A further
reason for this omission is the fact that the characterization of
the structure on larger scales seems to be a daunting task, since already the
classification of the local structure is highly complex.  However,
whether or not disordered systems have indeed a structural order that
extents beyond a few particle diameters is an important question since
it is, e.g., related to the formation of the critical nucleus for
crystallization or the possible growth of a static length scale that
is often invoked in rationalizing the slow dynamics in glass-forming
systems~\cite{kelton_10,binder_11,royall_17,rfot,adam_65,chandler_10}.
In the present work we use a novel approach to reveal that liquids do
have a highly non-trivial 3D structure that is surprisingly simple at length
scales beyond the first few neighbors.

In order to show the generality of our results we will consider two
systems that have a very different local structure: A binary mixture of
Lennard-Jones particles (BLJM), with 80\% A particles and 20\% B
particles~\cite{kob_95}, and silica (see Methods). The former liquid has a 
closed-packed local structure that is similar to the one of a hard sphere
system while the latter is a paradigm for an open network liquid with
local tetrahedral symmetry~\cite{binder_11}.

\begin{figure}[ht]
\center
\includegraphics[width=0.27\columnwidth]{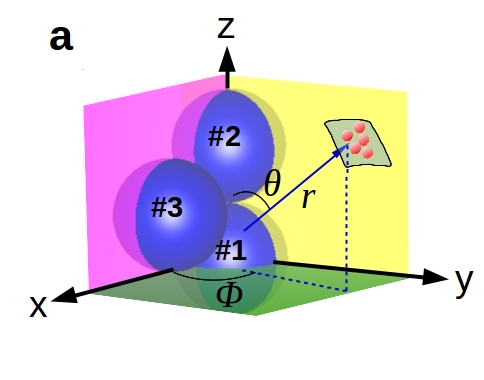}
\includegraphics[width=0.25\columnwidth]{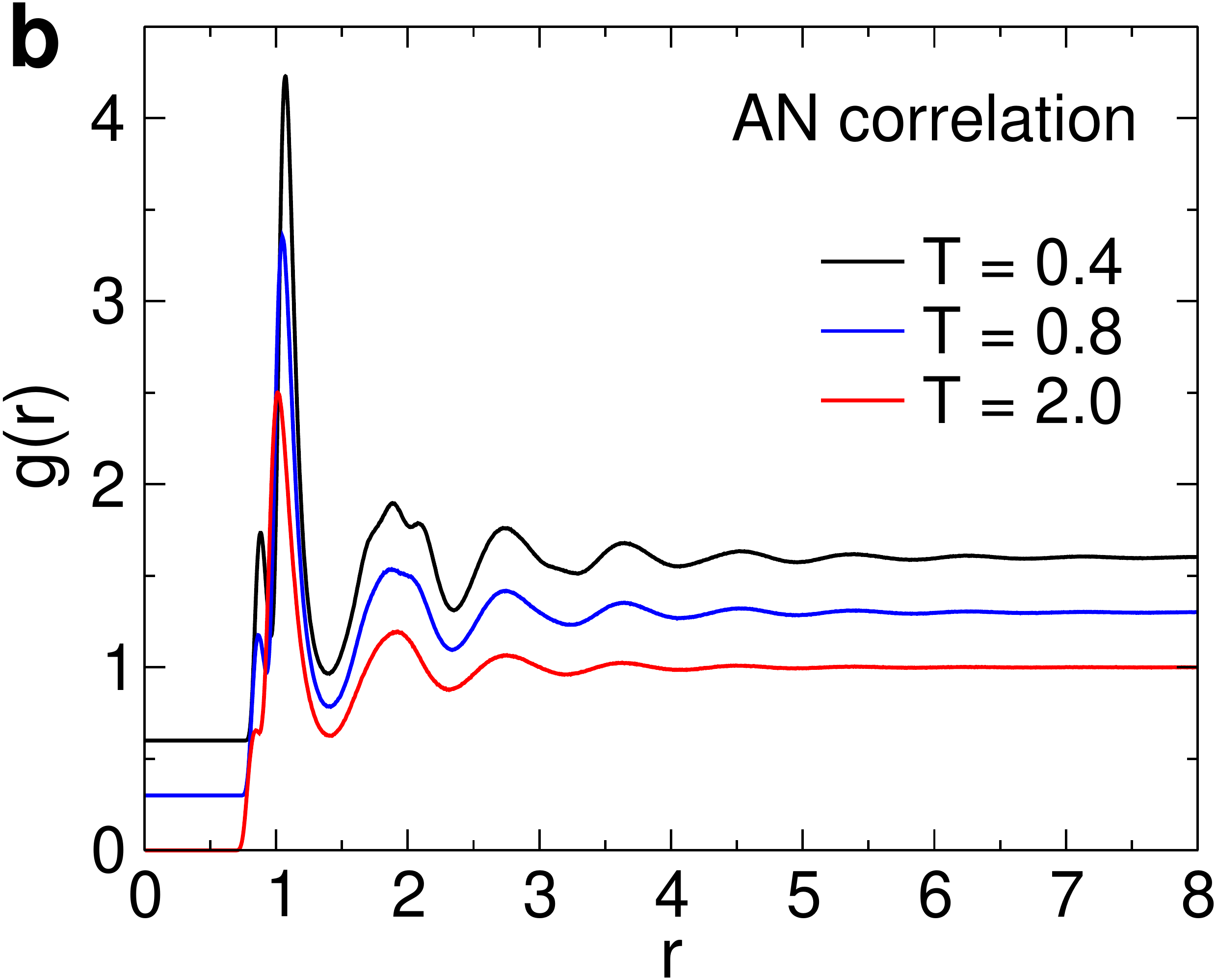}
\includegraphics[width=0.2\columnwidth]{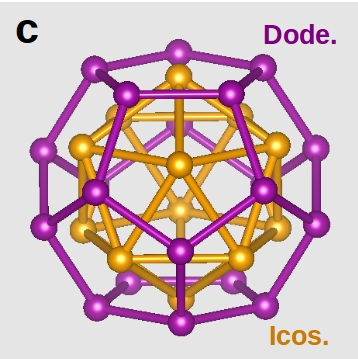}
\includegraphics[width=0.75\columnwidth]{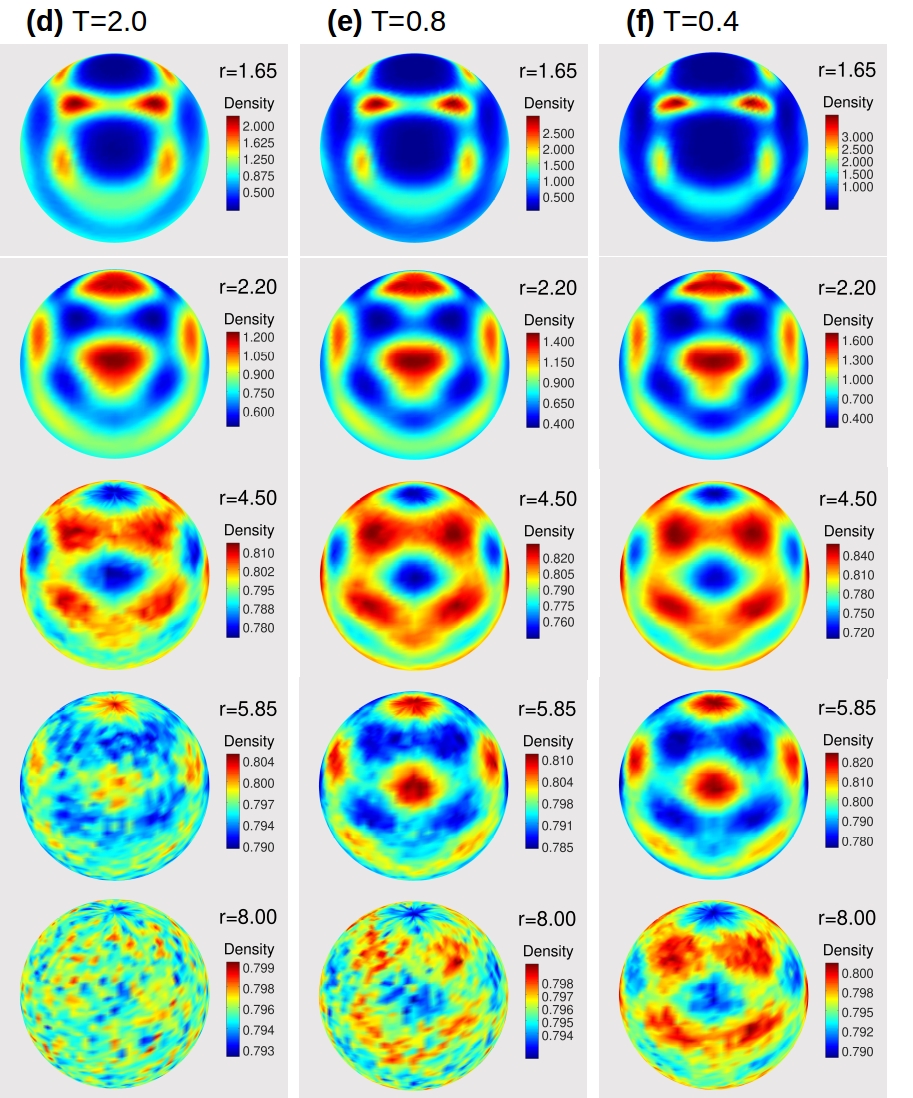}
\caption{
{\bf Distribution of particles in three dimensions for the BLJM.}
{\bf a}: The definition of the local coordinate system involves
three particles that are nearest neighbors to each other. 
{\bf b}:
Radial distribution function $g_{\rm AN}(r)$ for different temperatures (N=A+B). For the sake of clarity
the different curves have been shifted vertically by multiples of 0.6.
{\bf c}:
An icosahedron is the dual polyhedron of a dodecahedron and vice
versa. 
{\bf d} to {\bf f}: Density distribution $\rho(\theta,\phi,r$)
for different values of $r$, i.e., the distribution of the particles
that are in a spherical shell of radius $r$ and thickness 0.4
around the central particle. {\bf d}: $T=2.0$, {\bf e}: $T=0.8$, {\bf f}: $T=0.4$.
}
\label{fig1_densityplot_bljm}
\end{figure}

We study the equilibrium properties of the BLJM in a temperature range
in which the system changes from a very fluid state to a moderately
viscous one, i.e.~$5.0\geq T \geq 0.40$~\cite{kob_95}. Silica is studied at 3000~K,
a temperature at which the liquid is relatively viscous~\cite{horbach99}. To probe
the three dimensional structure of the BLJM we introduce a local
coordinate system as follows (Fig.~\ref{fig1_densityplot_bljm}{\bf a}):
Take any three A particles that touch each other, i.e., they form
a triangle with sides that are less than the location of the first
minimum in the radial distribution function $g(r)$, i.e.$\approx 1.4$
(see~Fig.~\ref{fig1_densityplot_bljm}{\bf b}). We define the position of particle \#1
as the origin, the direction from particle \#1 to \#2 as the $z-$axis,
and the plane containing the three particles as the $z-x-$plane. (For
the case of SiO$_2$ we use a Si atom as the central particle and
two nearest neighbor oxygen atoms to define the three coordinate
axes.) This local reference frame allows to introduce a spherical
coordinate system $\theta,\phi,r$ and to measure the probability of
finding any other particle at a given point in space, i.e.~to measure a
four point correlation function.  Note that this coordinate system can
be defined for all triplets of neighboring particles and these density
distributions can be averaged to improve the statistics. Since this
coordinate system is adapted to the configuration by the three particles,
it allows to detect angular correlations that are not visible in $g(r)$
or in previously considered structural observables.

\clearpage

For the BLJM Fig.~\ref{fig1_densityplot_bljm}{\bf d} shows the three dimensional
normalized distribution $\rho(\theta,\phi,r)$ of the particles on the
sphere of radius $r$ centered at an A particle. The temperature is
$T=2.0$, i.e. above the melting point of the system which is around $T=
1.0$~\cite{pedersen18}.  We recognize that $\rho(\theta,\phi,r)$ has a
noticeable angular dependence not only at small distances but also at
intermediate ones, i.e.~$r=4.5$, which corresponds to the fifth nearest
neighbor shell in $g(r)$ (panel {\bf b}). (Here we denote by $g(r)$ the
partial radial distribution function $g_{\rm AN}(r)$, where N stands
for A+B, see also~\ref{SI_fig_structure}.) If temperature is decreased to
$T=0.8$, panel {\bf e}, the angular signal can be easily detected up
to $r=5.9$ and for $T=0.40$, panel {\bf f}, even at $r=8.0$, i.e.~the
9th nearest neighbor shell. These snapshots show that this liquid has
a non-trivial angular correlation that extends to distances well beyond
the first few nearest neighbor shells. See Movies~1 and 2 for a dynamical
presentation of these results.

\begin{figure}[t]
\center
\includegraphics[width=0.99\columnwidth]{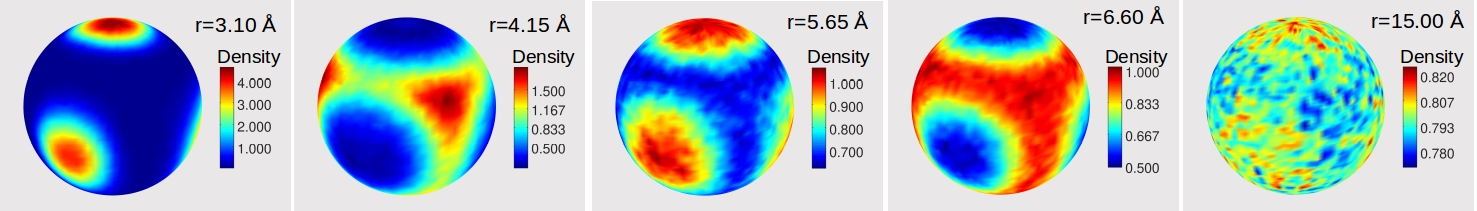}
\caption{
{\bf Distribution of particles in three dimensions for silica.}
$T=3000$~K. The thickness of the shell is 1~\AA. Depending on $r$,
the high/low density regions show a tetrahedral symmetry with two different symmetry axes.
}
\label{fig2_densityplot_sio2}
\end{figure}

Furthermore one notes that $\rho(\theta,\phi,r)$ has a highly symmetric
shape: For $r=1.65$, corresponding to the distance between the first
minimum and the second nearest neighbor peak in $g(r)$, one observes a
dodecahedral-like symmetry (see Fig.~\ref{fig1_densityplot_bljm}{\bf c}). For $r=2.2$ the distribution has
instead an icosahedral symmetry.  This result can be understood
by recalling that an icosahedron is the dual of a dodecahedron, and vice
versa (Fig.~\ref{fig1_densityplot_bljm}{\bf c}) and hence the local dips
formed by particles in the first minimum will be occupied by the particles
in the subsequent shell. As shown below, this ``duality mechanism'' works
even at large distances thus leading to a non-trivial angular correlation
in which as a function of $r$ density distributions with icosahedral
symmetry alternates with distributions with dodecahedral symmetry. Figures
\ref{fig1_densityplot_bljm}{\bf e} and {\bf f} show that with decreasing
temperature the intensity of the signal at intermediate and large
distances increases, indicating an enhanced order at low $T$.

Also for the case of the network liquid SiO$_2$ one finds
a pronounced anisotropy of the density correlation function,
Fig.~\ref{fig2_densityplot_sio2}. In contrast to the hard-sphere-like
liquid one finds here that the spherical shells with a pronounced
orientational order all show a tetrahedral symmetry which makes sense
since the dual of a tetrahedron is again a tetrahedron.  We emphasize
that for geometrical reasons at large $r$ a region with high
$\rho(\theta,\phi,r)$ is {\it not} a single particle, but a structure
that grows linearly with $r$ and hence is a whole collection of particles,
i.e., for fixed $r$ the structure is given by patches with a high density
of particles that alternate with patches with low density.

To analyze these findings in a quantitative manner we use the
standard procedure to decompose the signal on the sphere into
spherical harmonics $Y_l^m$, $\rho(\theta,\phi,r) =$\linebreak
$\sum_{l=0}^\infty \sum_{m=-l}^{l}\rho_l^m(r) Y_l^m(\theta,\phi)$,
where the expansion coefficients $\rho_l^m$ are given in the Methods,
and to consider the square root of the angular power spectrum
$S_\rho(l,r)= [(2l+1)^{-1}\sum_{m=-l}^{l}|\rho_l^m(r)|^2]^{1/2}$. For
the BLJM the component with $l=6$ has the largest amplitude
(\ref{SI_fig_sff_S_compare-l}), independent of $r$, a result that is
reasonable in view of the icosahedral and dodecahedral symmetries that
we find in the density distribution. For SiO$_2$ it is the component
$l=3$ which has the strongest signal since this mode captures well the
tetrahedral symmetry of the density field.

\begin{figure}[t]
\center
\includegraphics[width=0.48\columnwidth]{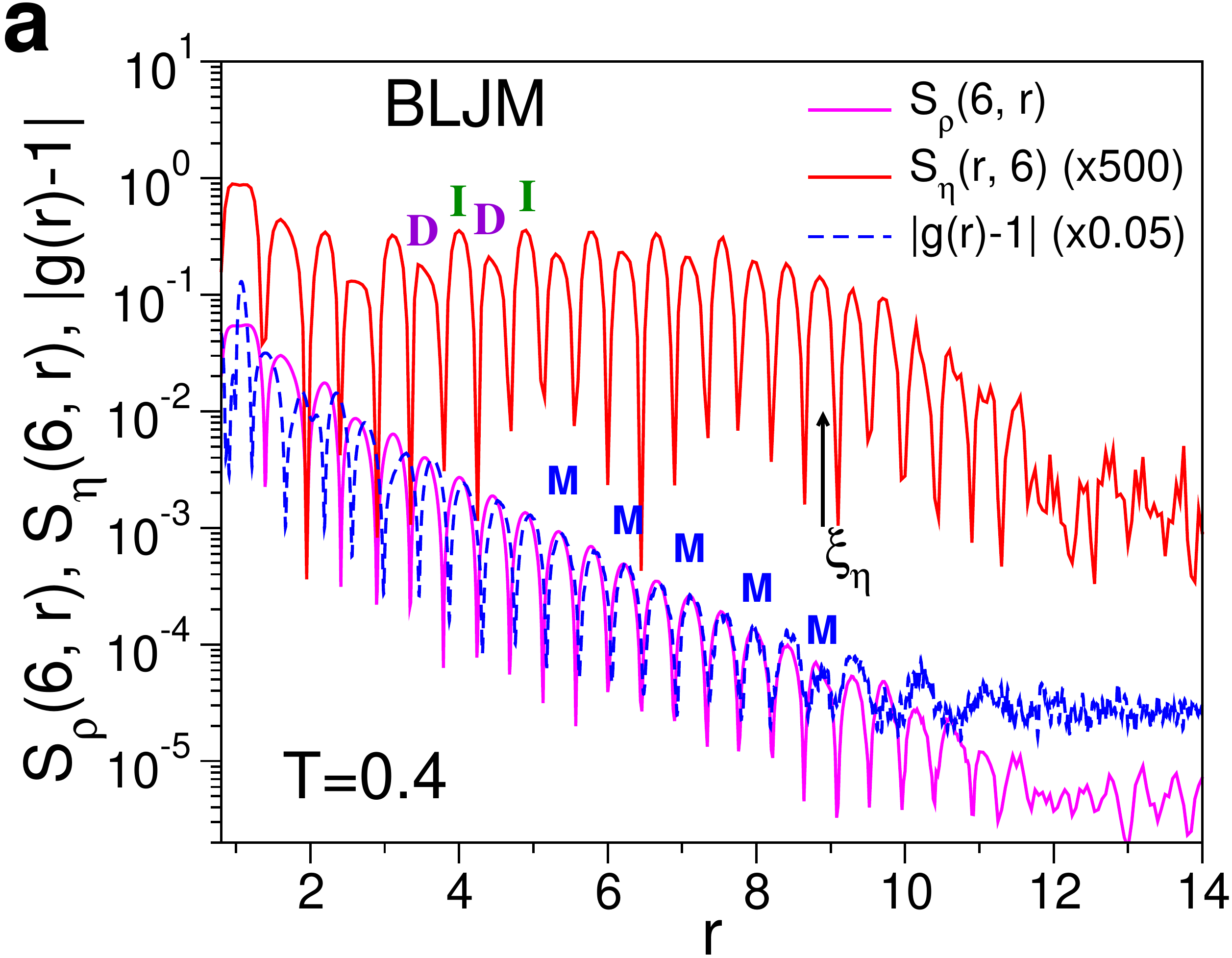}
\includegraphics[width=0.48\columnwidth]{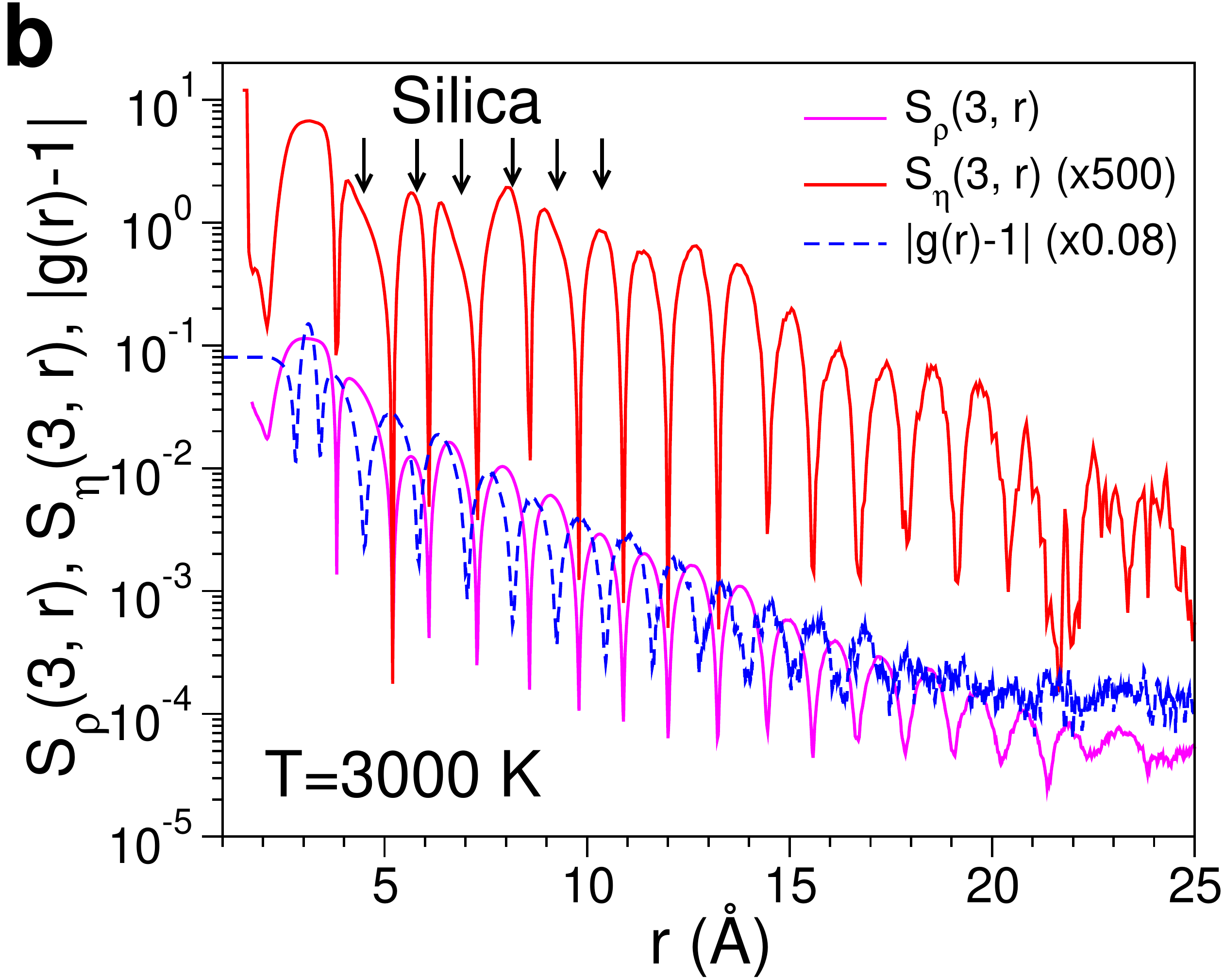}
\caption{
{\bf Quantitative characterization of the structural order.} {\bf a}
and {\bf b}: The angular power spectra and radial distribution function
for the BLJM at $T=0.4$, {\bf a}, and for silica at $T=3000$~K, {\bf b}.
The power spectra $S_\rho(6,r)$  and $S_\rho(3,r)$ (magenta curves)
show an exponential-like decay as a function of the distance $r$. The
power spectra for the normalized density distribution, $S_\eta(6,r)$
and $S_\eta(3,r)$ (red curves), stay large even at intermediate $r$.
For the BLJM and $r \gtrsim 4.0$ the high/low maxima in $S_\eta(r)$,
labeled I and D, coincide with the minima/maxima (labeled M) in $|g(r)-1|$
(blue line). This up-down behavior is related to the
alternating icosahedral/dodecahedral symmetry in the distribution of
the particles when $r$ is increased. For SiO$_2$ the arrows indicate the
distances at which $g_{\rm SiSi}(r)=1$. For better visibility $S_\eta(r)$
and $|g(r)-1|$ have been shifted vertically.
}
\label{fig3_srho_gr}
\end{figure}

In Fig.~\ref{fig3_srho_gr} we show the $r-$dependence of $S_\rho(l,r)$
and one sees that for both systems the signal
decays quickly with increasing $r$. Figure~\ref{fig1_densityplot_bljm}{\bf
f} shows that at distance $r=5.85$ the density distribution has a
pronounced structure while from Fig.~\ref{fig3_srho_gr}{\bf a} one sees that at
this $r$ the absolute value of $S_{\rho}(l,r)$ is small. This smallness
is due to the fact that $S_{\rho}(l,r)$ is not only sensitive to the
angular dependence of the distribution, but also to the amplitude of the
signal. In order to probe whether or not the density distribution has a
pronounced symmetry it is therefore useful to consider a {\it normalized}
density distribution $\eta(\theta,\phi,r)=[\rho(\theta,\phi,r)-\rho_{\rm
min}(r)]/[\rho_{\rm max}(r)-\rho_{\rm min}(r)]$, where $\rho_{\rm
max}(r)$ and $\rho_{\rm min}(r)$ are the maximum and minimum
of $\rho(\theta,\phi,r)$, respectively (at fixed $r$). The square root of the angular power
spectrum of $\eta(\theta,\phi,r)$, $S_{\eta}(r)$, is included in
Fig.~\ref{fig3_srho_gr}{\bf a} as well. We see that for the BLJM
$S_\eta$ oscillates around a constant value which demonstrates that
for this system the density distribution has a pronounced orientational
order even at large distances. For distances larger than
a threshold $\xi_{\eta}(T)$, $S_{\eta}(r)$ starts to decay before
it reaches at large $r$ a value that is determined by the noise of
the data. (See S.I. for a precise definition of $\xi_\eta$ and its
$T-$dependence.) For distances smaller than 2-3 particle diameters there
is no direct correlation between $S_{\rho}(r)$ and $g(r)$ since at these
$r$'s the local packing is determined also by energetic considereations,
see S.I.

Most remarkable is the observation that for the case of the BLJM
the height of the local maxima in $S_\eta(r)$ shows a periodic
behavior in that a high maximum is followed by a low one. A visual
inspection of $\rho(\theta,\phi,r)$ reveals that these high/low maxima
correspond to distances at which the distribution has a pronounced
icosahedral/dodecahedral symmetry demonstrating that these two Platonic
bodies are not only present at short distances but also at large ones,
in agreement with the snapshots in Fig.~\ref{fig1_densityplot_bljm}.
One thus concludes that for hard-sphere-like systems the distribution
of the particles in three dimensions is given by shells in which
particles are arranged in an icosahedral pattern, followed by a shell
in which there is a dodecahedral pattern. For distances larger than
$r\approx 4$ one finds that the radial positions of these two geometrical
arrangements match perfectly the locations of the minima/maxima in $g(r)$,
Fig.~\ref{fig3_srho_gr}{\bf a}. This observation can be rationalized
by the fact that a dodecahedron has 20 vertices (i.e., regions in
which $\rho(\theta,\phi,r)$ has high values) and an icosahedron only
12, thus making that the former structure corresponds to the {\it
maxima} of $g(r)$ and the latter to the {\it minima}.

In contrast to the BLJM we find that for silica,
Fig.~\ref{fig3_srho_gr}{\bf b}, the locations of the maxima in
$S_\rho(3,r)$ do not correspond to the ones in $|g_{\rm SiSi}(r)-1|$
but are instead close to distances at which $g(r)=g_{\rm SiSi}(r)=1$ (as
indicated by the arrows in the graph), i.e.~corresponds to a distance at which
one expects {\it no} structural order.  (See~\ref{SI_fig_structure}{\bf d}
for the Si-O partial radial correlation function.) This shows that for
liquids which have an open network structure, the distances at which one
finds the highest orientational symmetry is {\it not} associated with
a dense packing of particles, in contrast to hard-sphere-like systems.
Finally we note that for both systems the decay of $S_\rho(r)$ matches
very well the one of $g(r)$. This indicates that the two functions are
closely related to each other, i.e.~the loss in the symmetry of the
density field in three dimensions leads to the decay in the structure
as measured by $g(r)$, a result that is reasonable since the angular
integral of $\rho(\theta,\phi,r)$ is proportional to $g(r)$.

Figure~\ref{fig3_srho_gr}{\bf b} shows that also for silica
$S_{\eta}(r)$ is high for small and intermediate distances, but even
in this range it decreases slowly, indicating that for this network
liquid the orientational symmetry is gradually lost with increasing $r$.
This result might be related to the fact that the connectivity of the silica
network is lower than one of the densely packed hard-sphere-like liquid, hence
the former structure is more flexible and therefore it is more difficult to
propagate the orientational order in space to large distances.

\begin{figure}[ht]
\center
\includegraphics[width=0.47\columnwidth]{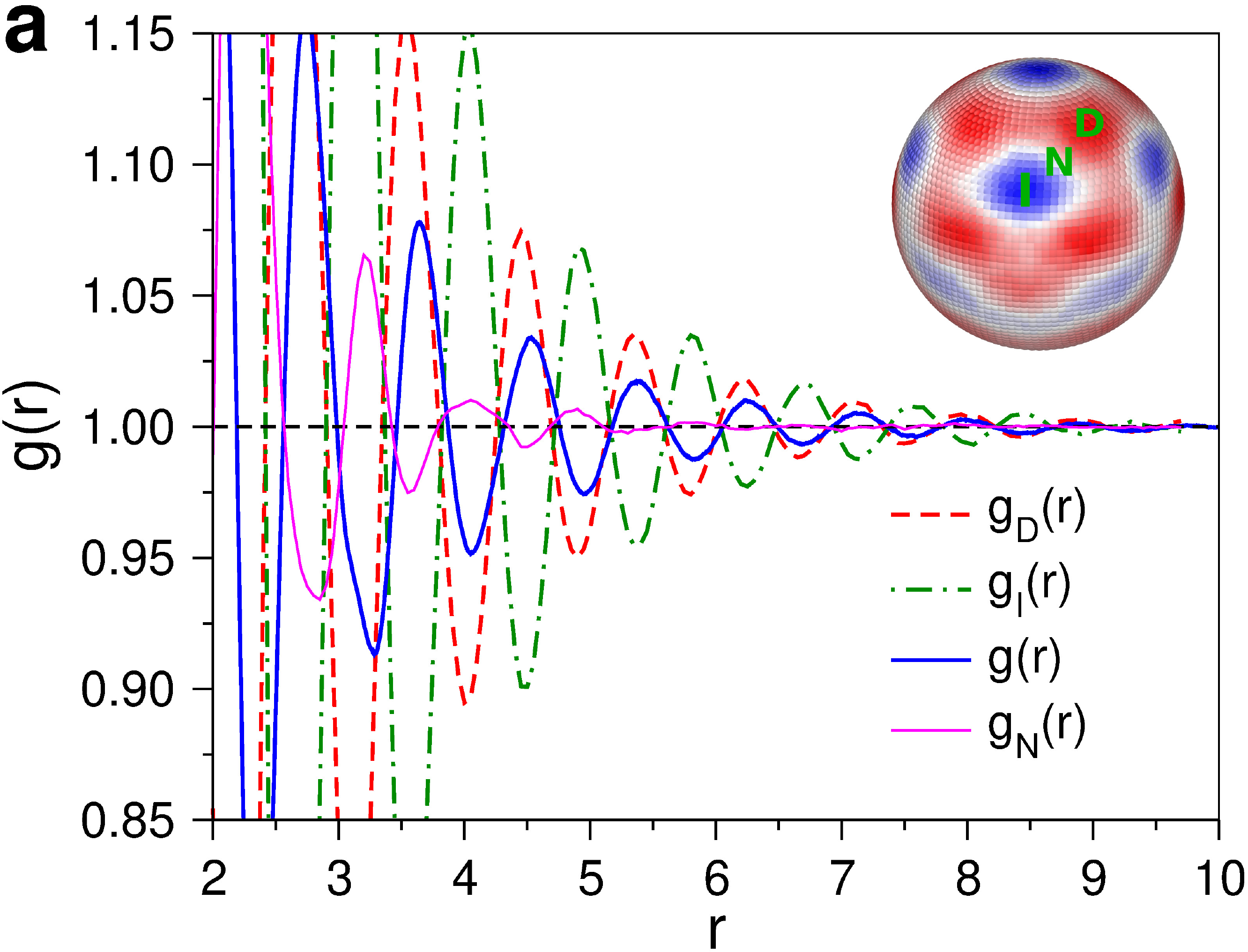}
\includegraphics[width=0.47\columnwidth]{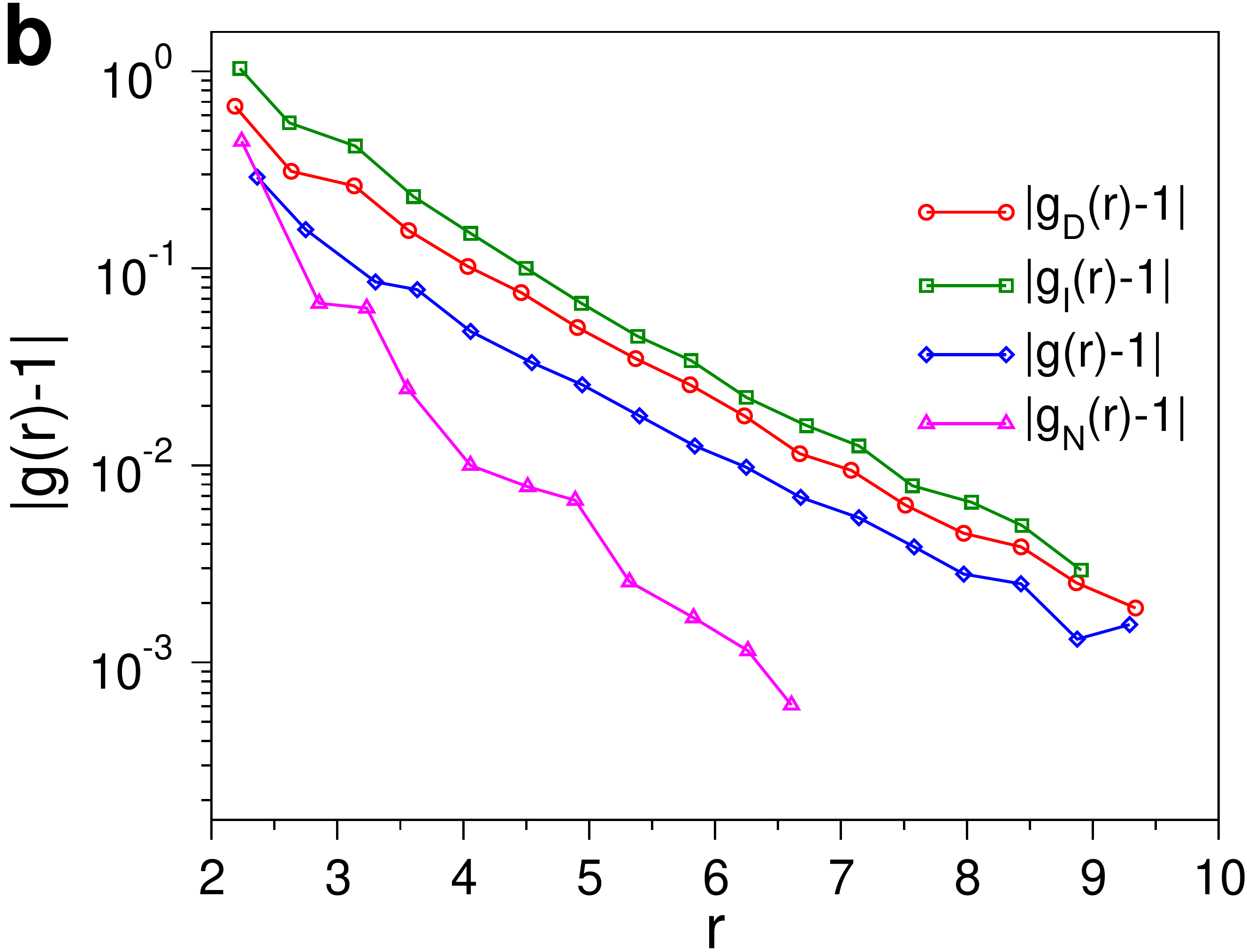}
\includegraphics[width=0.94\columnwidth]{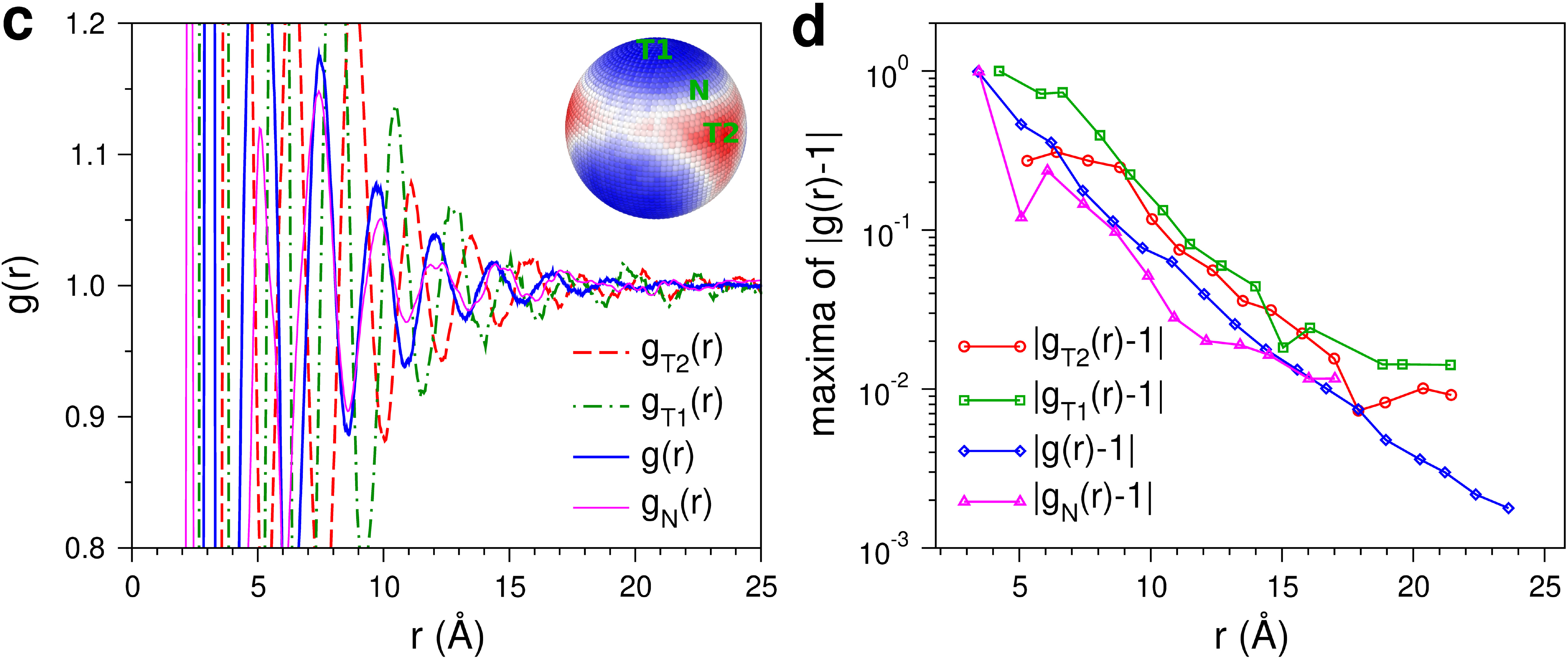}
\caption{{\bf Anisotropic radial distribution functions.}
Definition of the directions for measuring the radial distribution
functions:
{\bf a}: BLJM. Inset: I/D are the directions defined by the vertices of the
icosahedra/dodecahedra. N is the midpoint between these two directions. Main figure:
Radial distribution function as measured in the direction I, D, and N.
{\bf b}: Same quantities as in {\bf a} on logarithmic scale. For the sake of clarity
only the maxima in the curves are shown.
{\bf c}: Silica. Inset:
T1 and T2 are the directions defined by the two interlocked tetrahedra.
N is the midpoint between these two directions.
Main figure: Radial distribution function as measured in the direction T1, T2, and N.
{\bf d}: Same quantities as in {\bf c} on logarithmic scale. For the sake of clarity
only the maxima in the curves are shown.
}
\label{fig5_gr_anisotropic}
\end{figure}

Since we have found that the distribution of the particles around a
central particle is aniso\-tropic, it is of interest to consider also the
radial distribution functions in which one probes the correlations in a
specific direction with respect to the local coordinate system shown in
Fig.~\ref{fig1_densityplot_bljm}{\bf a}. This type of information can
be obtained for colloidal systems from confocal microscopy experiments
and, more indirectly, from scattering experiments~\cite{wochner_09}.
The insets in Fig.~\ref{fig5_gr_anisotropic}{\bf a}/{\bf c} show the directions we considered for
the two type of liquids: For the BLJM the directions defined by the vertices
of the icosahedra/dodecahedra, and the directions given by the midpoints between
these two type of vertices and for silica the directions of the vertices
of the tetrahedra, the points given by the midpoints of the faces,
and the directions given by the midpoints between the two former directions.

In Fig.~\ref{fig5_gr_anisotropic}{\bf a} we show for the BLJM the radial
distribution functions for these special directions and one recognizes
that the amplitude of the signal depends indeed strongly on the direction
considered. For the directions of the icosahedra and of the dodecahedra,
$g_I(r)$ and $g_D(r)$, respectively, we find for intermediate and
large distances, that $g_D(r)$ oscillates in phase with
$g(r)$ whereas $g_I(r)$ has oscillations that are in anti-phase. These
observations are coherent with the aforementioned argument that the
number of vertices in the dodecahedra exceed the ones for the icosahedra.

\clearpage

Furthermore Fig.~\ref{fig5_gr_anisotropic}{\bf a} shows that the
amplitudes of the oscillations in $g_I(r)$ and $g_D(r)$ are significantly
larger than the ones found in $g(r)$, a result that is reasonable
since the latter function is a weighted average of the two former ones
and hence will be affected by cancellation effects. The distribution
function in the direction that corresponds to the mid-point of the line
connecting two neighboring vertices of an icosahedron and a dodecahedron,
$g_N(r)$, shows significantly smaller oscillations than $g(r)$, a result
that is expected since one probes the structure in a direction which
does not pass close to the locations that correspond to the vertices
of the icosahedra/dodecahedra. Figure~\ref{fig5_gr_anisotropic}{\bf b}
shows these distribution functions on a logarithmic scale. (For the sake
of clarity only the maxima and minima of the functions are shown.) One
notices that the slope of the curves for $g(r)$, $g_I(r)$, and $g_D(r)$, i.e.~the length
scale over which the correlation decays, is basically independent of the
function considered, demonstrating that they are indeed closely related
to each other. In contrast to this the data from $g_N(r)$ decay faster
showing that in this direction the correlation length is smaller.

For the case of silica the connection between the extrema in $g(r)$
with the ones obtained from the radial distribution functions in the
special directions T1 and T2 (see inset Fig.~\ref{fig5_gr_anisotropic}{\bf
c}) is not straightforward. One finds that the peaks in $g_{\rm
T1}(r)$ ($g_{\rm T2}$) are where $g(r)$ rises (decreases) quickly,
Fig.~\ref{fig5_gr_anisotropic}{\bf e}. In fact the extrema of $g_{\rm
T1}(r)$ are very close to the distance at which $g(r)$ becomes 1.0,
i.e. an $r$ at which the Si density corresponds to the one expected for
an ideal gas. The reason for this is presently not known and thus it
will be interesting to determine whether this is a general feature for
liquids that have an open network structure.

\begin{figure}[th]
\center
\includegraphics[width=0.50\columnwidth]{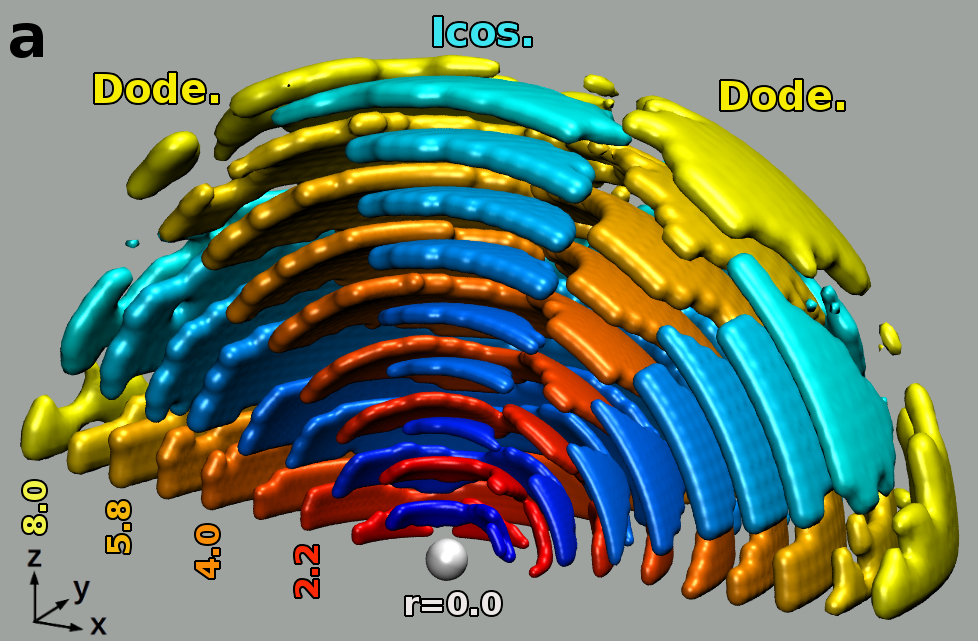}
\includegraphics[width=0.374\columnwidth]{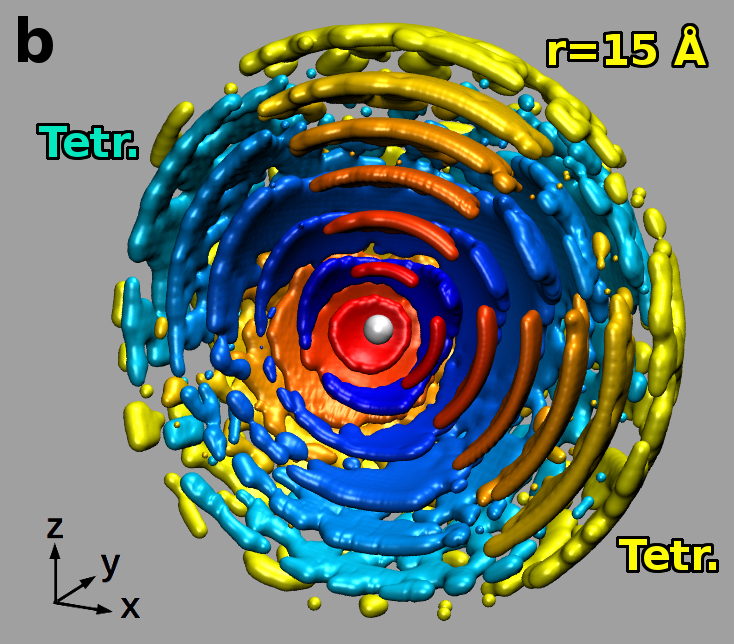}
\caption{
{\bf Three dimensional representation of the density field.}
The shown layers correspond to distances at which $S_\rho(r)$ has a
local maximum.  Only regions with high density (covering 35\% area of
the sphere) are depicted. 
{\bf a}: BLJM at $T=0.4$. The bluish/reddish
colors correspond to the locations of the high/low maxima in $S_\eta(r)$
and thus to shells with icosahedral/dodecahedral symmetry.
{\bf b}: Silica at $T=3000$K. The high density regions form interlocked
zones with a tetrahedral symmetry.
}
\label{fig6_3dstructure}
\end{figure}

For the case of silica the connection between the extrema in $g(r)$
with the ones obtained from the radial distribution functions in the
special directions T1 and T2 (see inset Fig.~\ref{fig5_gr_anisotropic}{\bf
c}) is not straightforward. One finds that the peaks in $g_{\rm
T1}(r)$ ($g_{\rm T2}$) are where $g(r)$ rises (decreases) quickly,
Fig.~\ref{fig5_gr_anisotropic}{\bf e}. In fact the extrema of $g_{\rm
T1}(r)$ are very close to the distance at which $g(r)$ becomes 1.0,
i.e. an $r$ at which the Si density corresponds to the one expected for
an ideal gas. The reason for this is presently not known and thus it
will be interesting to determine whether this is a general feature for
liquids that have an open network structure.

The radial distribution function for the ``neutral'' direction $N$
has a signal that is in phase with $g(r)$ and its amplitude is smaller
than the one of $g(r)$. The latter result is expected since one measures
the density field in a direction in which the fluctuations between the
interlocked tetrahedra basically cancel each other. In panel {\bf d}
we show the same radial distribution functions in a log-lin plot (only
the maxima are shown) and one recognizes that all of them decay in the
same exponential manner with a slope that is independent of the direction.

To give a comprehensive view of the particle arrangement in three
dimensions we present in Fig.~\ref{fig6_3dstructure} the density
distribution of the two systems. The colored regions correspond to
the zones in which the particle density is high and, by construction,
they cover 35\% of the sphere.  For the BLJM at intermediate and large
distances one recognizes clearly the presence of high density zones
with icosahedra symmetry (bluish color) interlocked with zones with
dodecahedra symmetry (yellow). The directions in which the blue and
yellow regions touch each other correspond to the neutral direction N
defined above and in which the particle correlation is weak.
For silica one finds instead interlocked tetrahedra at all distances,
panel {\bf b}. Again, the neutral direction corresponds to the one in
which the blue and yellow regions touch.

In conclusion we have demonstrated that liquids have non-trivial
structural symmetries at intermediate distances that have gone unnoticed
so far.  This result has been obtained by using a novel method which takes
into account the three dimensional angular dependence of the structure
and which can be readily applied to any system for which the particle
coordinates are accessible, such as colloidal and granular systems, or
materials in which some of the particles have been marked by fluorescence
techniques~\cite{xia_17,kegel_00,weeks_00,sherson_10,kou_17}. Since we
find that the nature of the orientational order depends on the system
considered, the method allows to make a more precise classification of
the structure of liquids, an aspect that should trigger the improvement
of experimental techniques that probe this structural order.


\section*{Acknowledgments}
 We thank D. Coslovich, G. Monaco, M. Ozawa, and K. Schweizer for discussions. 
{\bf {Funding:}}
Part of this work was supported by the China Scholarship Council grant 201606050112 
and grant ANR-15-CE30-0003-02.

{\bf {Author contributions:}}
Z.Z. and W.K. designed the research and carried out the simulations. Z.Z. analyzed the data. Z.Z. and W.K. wrote the paper.
{\bf {Competing interests:}} The authors declare no competing financial interests. 
{\bf {Data and materials availability:}} All data in the manuscript or
the Materials are available from W. Kob upon reasonable request.

\clearpage

\section*{Methods}
{\bf System and simulations.} 
The BLJM we study is a 80:20 mixture of Lennard-Jones
particles (type A and B) with interactions
given by $V_{\alpha\beta}(r)=4\epsilon_{\alpha\beta}
[(\sigma_{\alpha\beta}/r)^{12}-(\sigma_{\alpha\beta}/r)^{6}]$, where
$\alpha,\beta \in \{A,B\}$, $\sigma_{AA}=1.0$, $\epsilon_{AA}=1.0$,
$\sigma_{AB}=0.8$, $\epsilon_{AB}=1.5$, $\sigma_{BB}=0.88$, and
$\epsilon_{BB}=0.5$~\cite{kob_95}. Here we use $\sigma_{AA}$ and
$\epsilon_{AA}$ as the units of length and energy, respectively. We set
the mass of all particles equal to $m=1.0$ and the Boltzmann constant
is $k_B=1.0$. We simulate a total of $10^5$ particles at constant
volume (box size 43.68) and temperature. At the lowest temperature,
$T=0.40$, the run was $1.4\cdot 10^8$ time steps (step size 0.005) for
equilibration and the same length for production, time spans that are
sufficiently large to completely equilibrate the system, i.e.~the mean
squared displacement (MSD) of the particles was larger than 1.0. For
higher $T$'s the MSD was significantly larger. For the analysis of
the data we used 8 and 20 configurations for $S_\rho$ and $g(r)$,
respectively. Previous studies have shown that this system starts to show
a tendency to crystallization at temperatures around 0.4 if the system is
simulated for several $\alpha-$relaxation times~$^{29}$.
However, our simulations lasted only a few
$\alpha-$relaxation times (the mean squared displacmenet has reached 1-2),
which is long enough for equilibrating the liquid but not long enough
for allowing the system to crystallize  (as indicated by
the static structure factor, see S.I.). Also, all of our results show
a completely smooth dependence on temperature and hence it is unlikely
that they are affected by the presence of crystalline order.

For the simulation of silica, we use a recently optimized interaction
potential called SHIK, which has been show to be able to
describe reliably the properties of real silica~$^{30}$. A cubic simulation box
containing 120000 atoms was used, which corresponds at room temperature
and zero pressure to a box size of about 120~\AA.  The simulation was
carried out in the $NPT$ ensemble at 3000~K for $10^6$ time steps (step
size 1.6~fs). After equilibration, i.e.~the MSD for the Si atoms reached
several \AA$^2$, we collected 8 configurations spaced by $10^5$ time steps for
the subsequent structural analysis. Note that the melting temperature of
silica is around 2000~K, thus our simulations are far above the melting
point and the system is in its stable liquid phase.

For both systems we used the LAMMPS software~$^{31}$ to carry out the
simulations.\\

\noindent
{\bf Angular power spectrum.}
The coefficient $\rho_l^m$ for the expansion of the density distribution into
spherical harmonics is given by
\begin{equation}
\rho_l^m =\int_0^{2\pi} d\phi \int_0^\pi d\theta \sin \theta  
\rho(\theta,\phi,r) Y_l^{m*}(\theta,\phi) \quad ,
\nonumber
\end{equation}

\noindent
where $Y_l^{m*}$ is the complex conjugate of the spherical harmonic
function of degree $l$ and order $m$. In practice this integration was
done for the BLJM by sampling the integrand over up to $2\cdot 10^9$ points for each
shell of width 0.4. The corresponding numbers for SiO$_2$ are 
$10^8$ points and 1.0~\AA. \\

\newpage
\setcounter{page}{1}
\centerline {\LARGE Supplemental Information}
\vspace{0.5cm}
{\Large {\bf Revealing the three-dimensional structure of liquids\newline
using four-point correlation functions}}\\
\vspace{1cm}
\centerline {\large Zhen Zhang, Walter Kob$^\ast$}
\centerline {\large $^\ast$Correspondence to:  walter.kob@umontpellier.fr}
\vspace{2cm}

\renewcommand{\figurename}{}
\renewcommand{\thefigure}{S.I. Fig. \arabic{figure}}
\setcounter{figure}{0}

\begin{figure}[ht]
\center
\includegraphics[width=0.45\columnwidth]{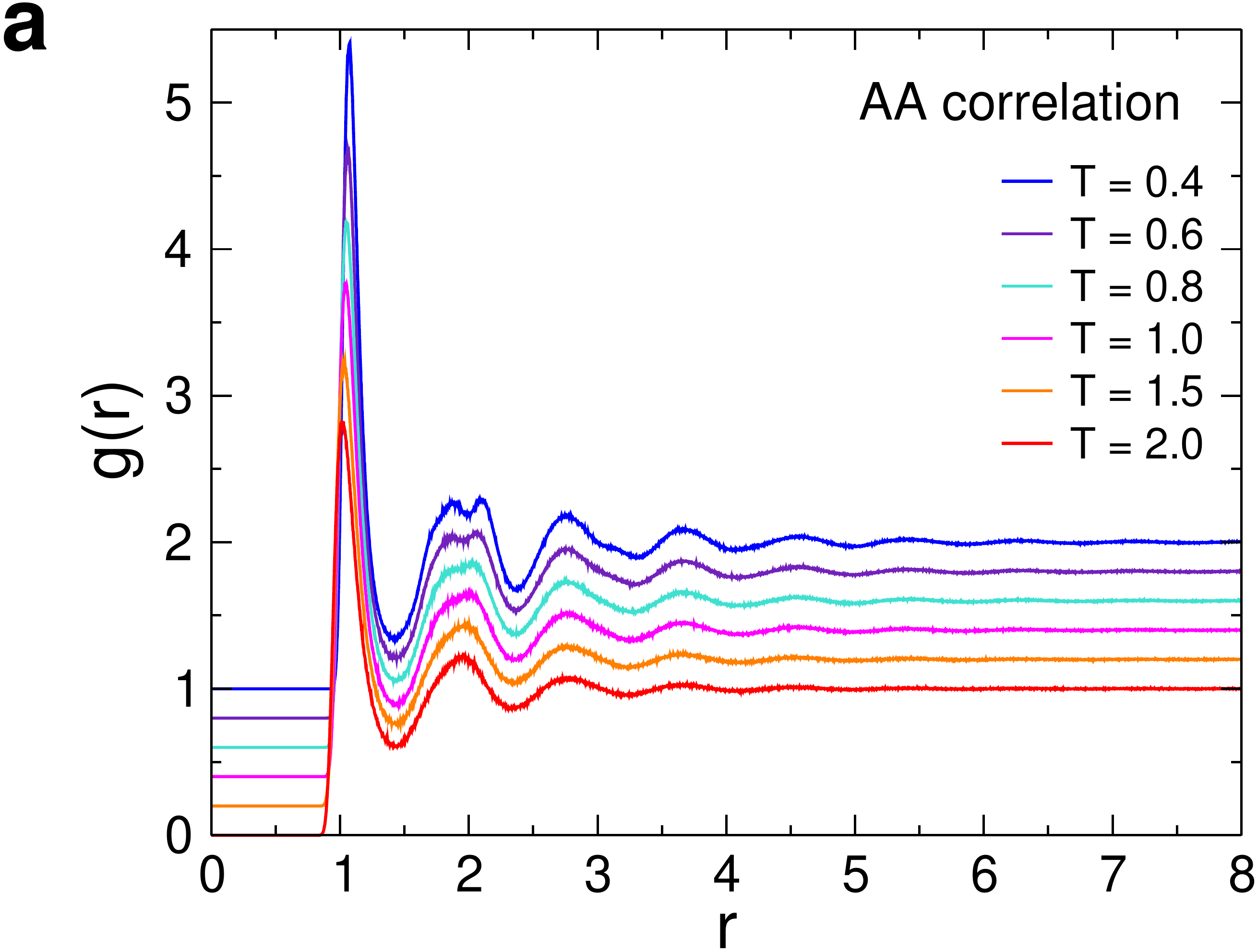}
\includegraphics[width=0.45\columnwidth]{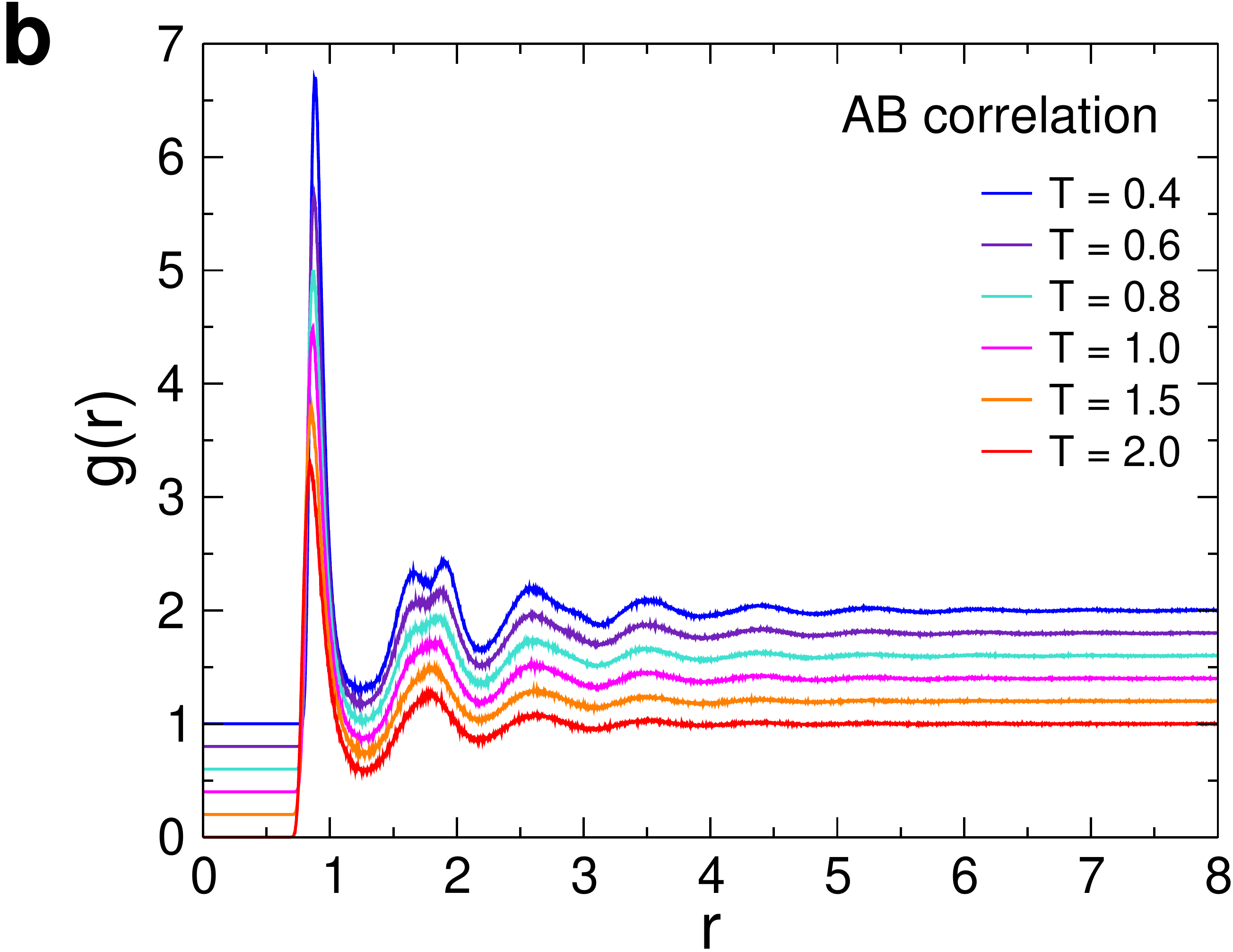}
\includegraphics[width=0.45\columnwidth]{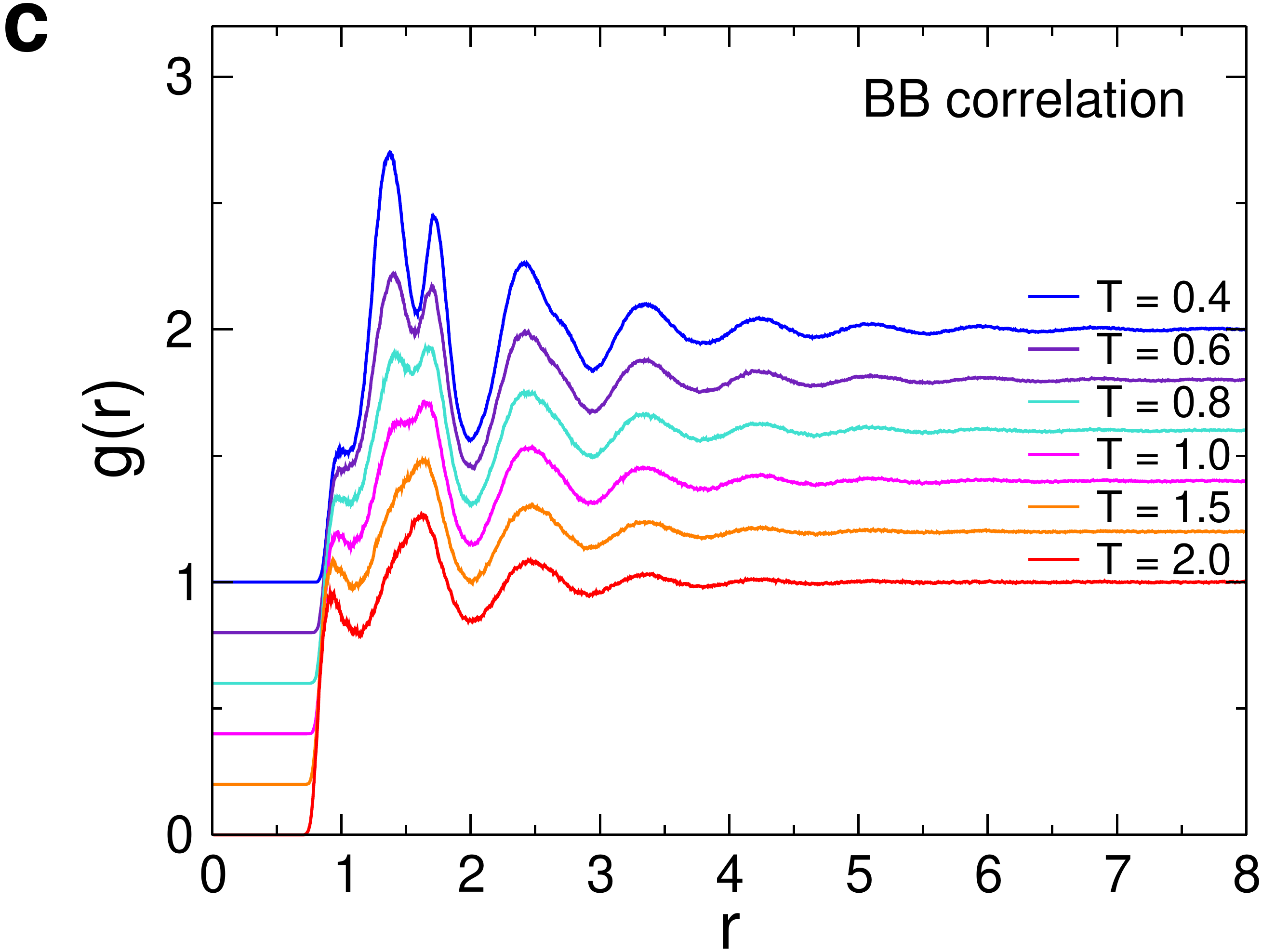}
\includegraphics[width=0.45\columnwidth]{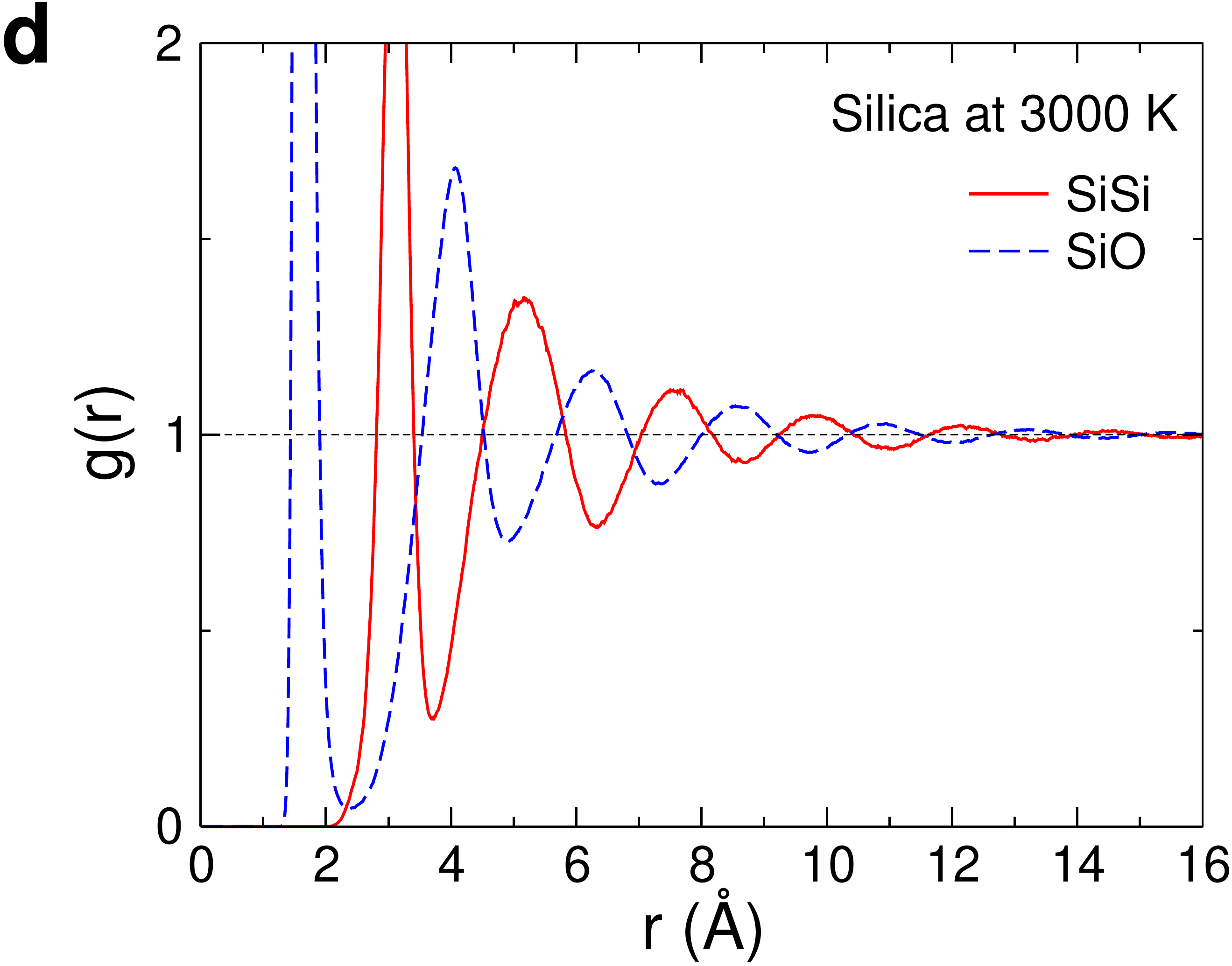}
\includegraphics[width=0.45\columnwidth]{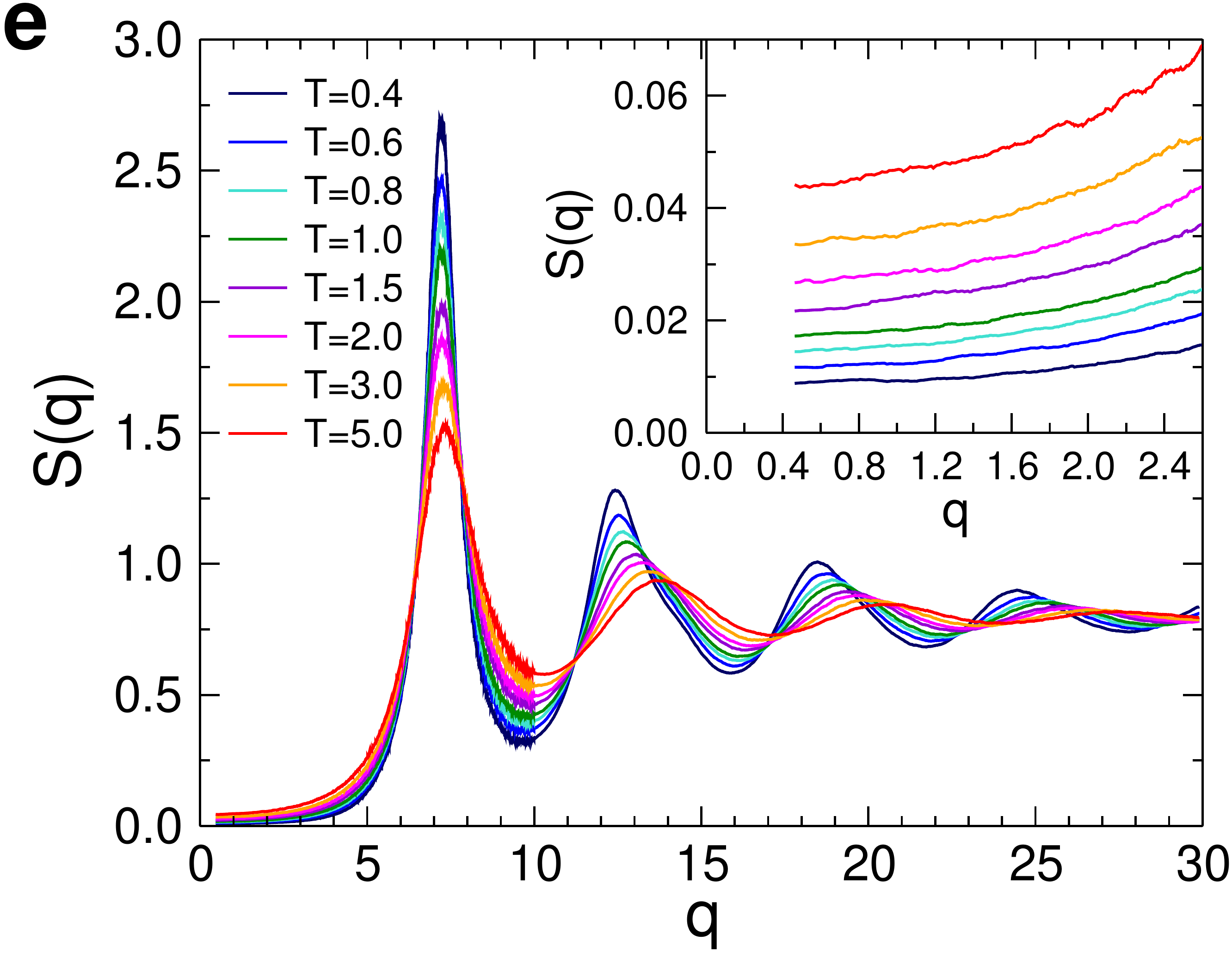}
\caption{{\bf Radial distribution functions for several temperatures.}
The AA, AB, and BB correlations for the BLJM are shown
in panels {\bf a}, {\bf b}, and {\bf c}, respectively. For the sake of clarity the different curves
have been shifted vertically by multiples of 0.2. 
{\bf d}: Partial radial distribution
functions for SiO$_2$ at $T=3000$~K.
{\bf e}: Partial static structure factor $S(q)$ for the AA pairs in the BLJM.
The inset shows $S(q)$ at small $q$.
}
\label{SI_fig_structure}
\end{figure}

\noindent
{\bf Radial distribution functions and static structure factor}\\
In \ref{SI_fig_structure} we show for the BLJM the three partial
radial distribution functions for different temperatures (see legend).
These graphs demonstrate that these functions show no marked peaks and
that their $T-$dependence is very smooth, as expected for a system that
is a good glass-former. Panel {\bf d} shows for SiO$_2$ at 3000~K the two
partial radial distribution functions related to the silicon atoms. Also
for this system we see that these two-point correlation functions show
no sharp peaks, i.e.~no indication for the presence of crystallites. \\

Since the presence of cystallization is easier to see in the reciprocal
space, we have for the BLJM also calculated the static structure factor $S(q)$.
This quantity was determined directly from the positions of the particles,
i.e.,\\[-5mm]

\begin{equation}
S({\vec q}) =\frac{1}{N}\sum_{j=1}^N \sum_{k=1}^N\exp[{i\vec q}\cdot ({\vec r}_j-{\vec r}_k)]
\quad .
\label{SI_eq_sq}
\end{equation}

\noindent
Since the system is isotropic, we have averaged $S({\vec q})$ over all
wave-vectors ${\vec q}$ that have the same norm $q=|{\vec q}|$. In
\ref{SI_fig_structure}{\bf e} we show the $q-$dependence of $S(q)$
for different temperatures. It can be seen that the $q-$dependence
of the structure is a smooth function of temperature. Also, the
curves show no signs for the growth of sharp peaks, also not at small
wave-vectors (see inset), which is further evidence that this system
does not crystallize even at the lowest temperature in the time window
we have probed. Qualitatively similar results are obtained for the case
of SiO$_2$.

\clearpage

\begin{figure}[ht]
\includegraphics[width=0.95\columnwidth]{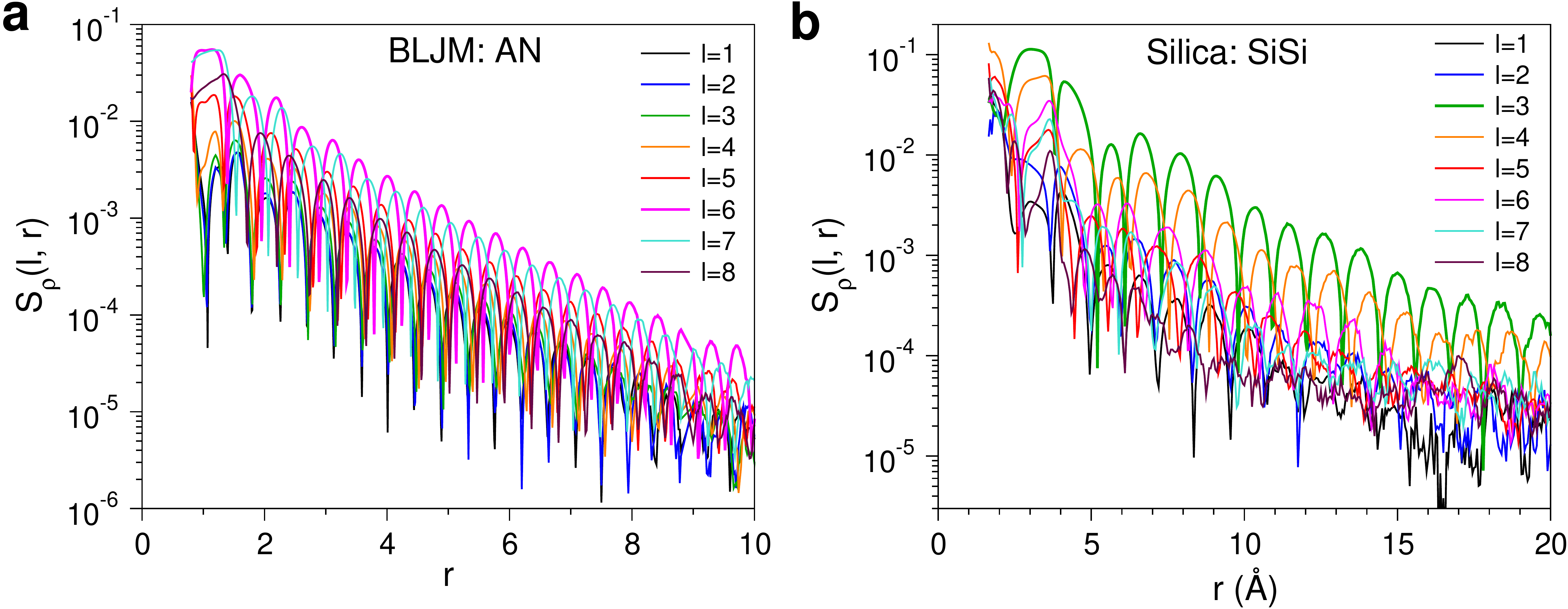}
\caption{{\bf $l-$dependence of the angular power spectrum.}
{\bf a:} $S_\rho(l,r)$ for the BLJM at $T=0.5$. The curves correspond to different values
of $l$. One sees that the maximum signal for $l=6$ is higher than the one of the other curves,
showing that this $l$ is the best choice to see the orientational order in the system.
{\bf b:} Same quantity for silica at 3000~K. Here it is the $l=3$ curve that shows the highest maxima.
}
\label{SI_fig_sff_S_compare-l}
\end{figure}

\noindent
{\bf $l-$dependence of the angular power spectrum}\\
In the main text we focus for the BLJM on the results for the index
$l=6$ in the expansion of the spherical harmonics of the density
distribution. In \ref{SI_fig_sff_S_compare-l}{\bf a} we show the
$r-$dependence of the angular power spectrum $S_\rho(l,r)$ for other
values of $l$. From this graph one recognizes that for $l=6$ the
signal dominates the other curves for most distances and hence this
value for the index is a good choice for probing the structural order
in the liquid. In \ref{SI_fig_sff_S_compare-l}{\bf b} we show the same
information for the case of silica and we see that here the curve for $l=3$
is the one with the highest maxima. From these figures it also becomes
clear that although the absolute height of the curves depends on $l$,
the general $r-$dependence is independent of $l$ and that each curve has
the same periodicity, for intermediate and large $r$.  This result shows
that the symmetry properties of the orientational order in the different
coordination shells is indeed independent of $r$ (if $r$ is not small).\\

\begin{figure}[th]
\center
\includegraphics[width=0.7\columnwidth]{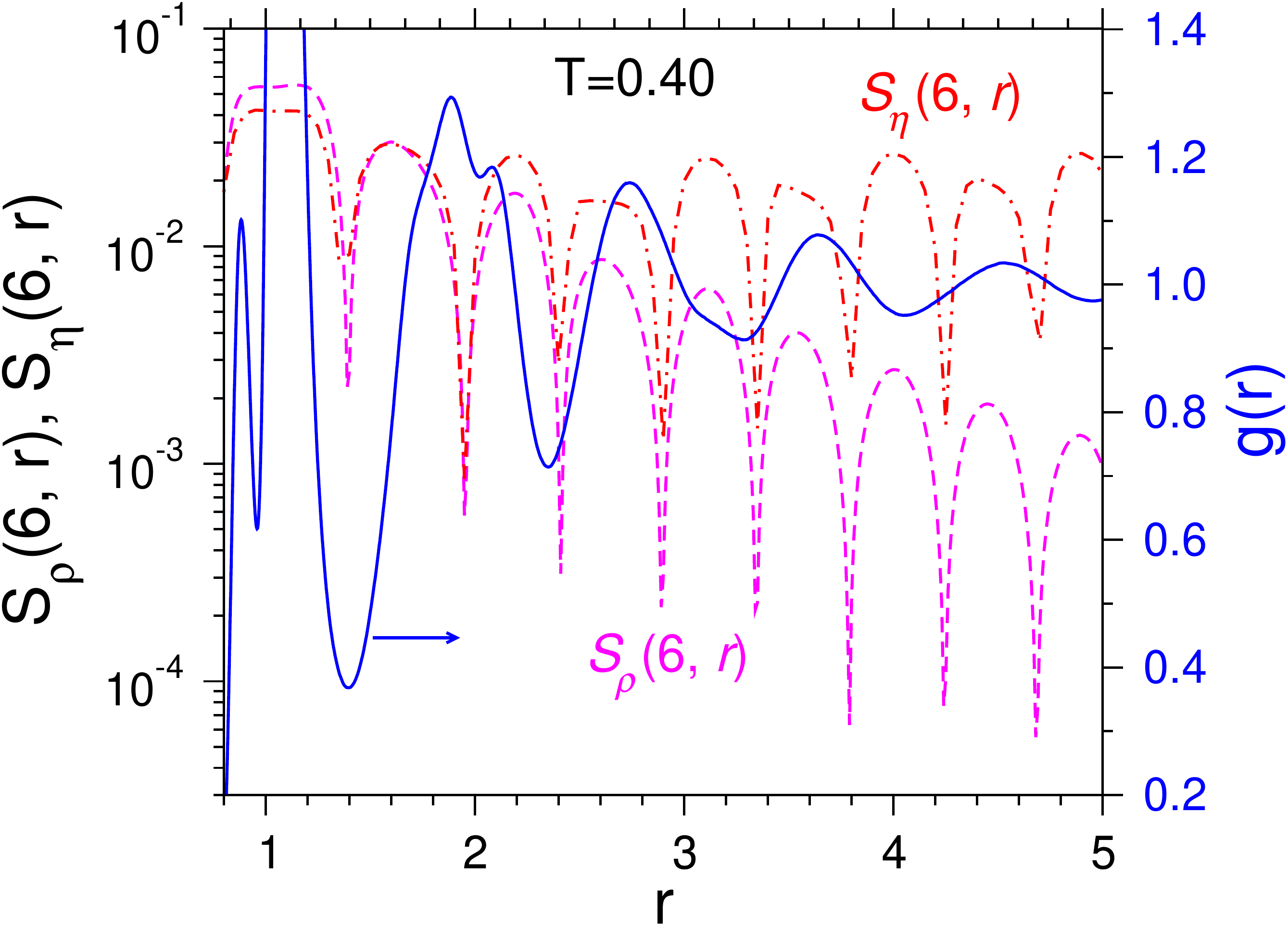}
\caption{{\bf Angular power spectra and radial distribution function at short distances for the
BLJM.}
$T = 0.4$ and $l=6$.  Note that the double peaks in the first shell, i.e. $r
\approx 1.0$ originate from A-B (smaller peak) and A-A (bigger peak)
correlations (see~\ref{SI_fig_structure}).
}
\label{SI_fig4_S_small_r}
\end{figure}

\noindent
{\bf Angular power spectra and radial distribution function at short distances}\\
In Fig.~\ref{fig3_srho_gr} of the main text we have shown that for
intermediate and large distances the angular power spectra $S_\rho(r)$
and $S_\eta(r)$ show oscillations with the same periodicity as the one
of the radial distribution function $g(r)$. In \ref{SI_fig4_S_small_r}
we show for the case of the BLJM these functions at small distances,
i.e., $r<5.0$. One recognizes from this graph that at these small
distances, in particular for $r<3.0$, the $r-$dependence of $g(r)$
is rather complex because of the local packing effects of the particles,
a behavior that is in agreement with previous studies of this and similar
systems~\cite{miracle_04,xia_17,royall_15,malins_13,dunleavy_15,coslovich_07,tanaka_12}.
Only for distances larger than around 4.0 the $r-$dependence of $g(r)$
becomes simpler. This behavior is also reflected in the $r-$dependence
of $S_\rho(r)$ and $S_\eta(r)$ that show for $ \leq 3.0$ a succesion of peaks of various
width and shapes signaling that at these short distances there is no
unique orientational symmetry. Only at larger distances do these functions
start to show a more regular behavior and they start to oscillate in phase
with $g(r)$.  Thus at the distance around $r=3$ we have a crossover
between a structure that is determined by local packing effects to a
structure that is determined by symmetry considerations.  This symmetry
at large $r$ is in turn determined by the packing at {\it small r},
i.e., in the case of the BLJM an icosahedra-like structure, while for
silia one has a tetrahedral symmetry.\\

\begin{figure}[ht]
{\center
\includegraphics[width=0.6\columnwidth]{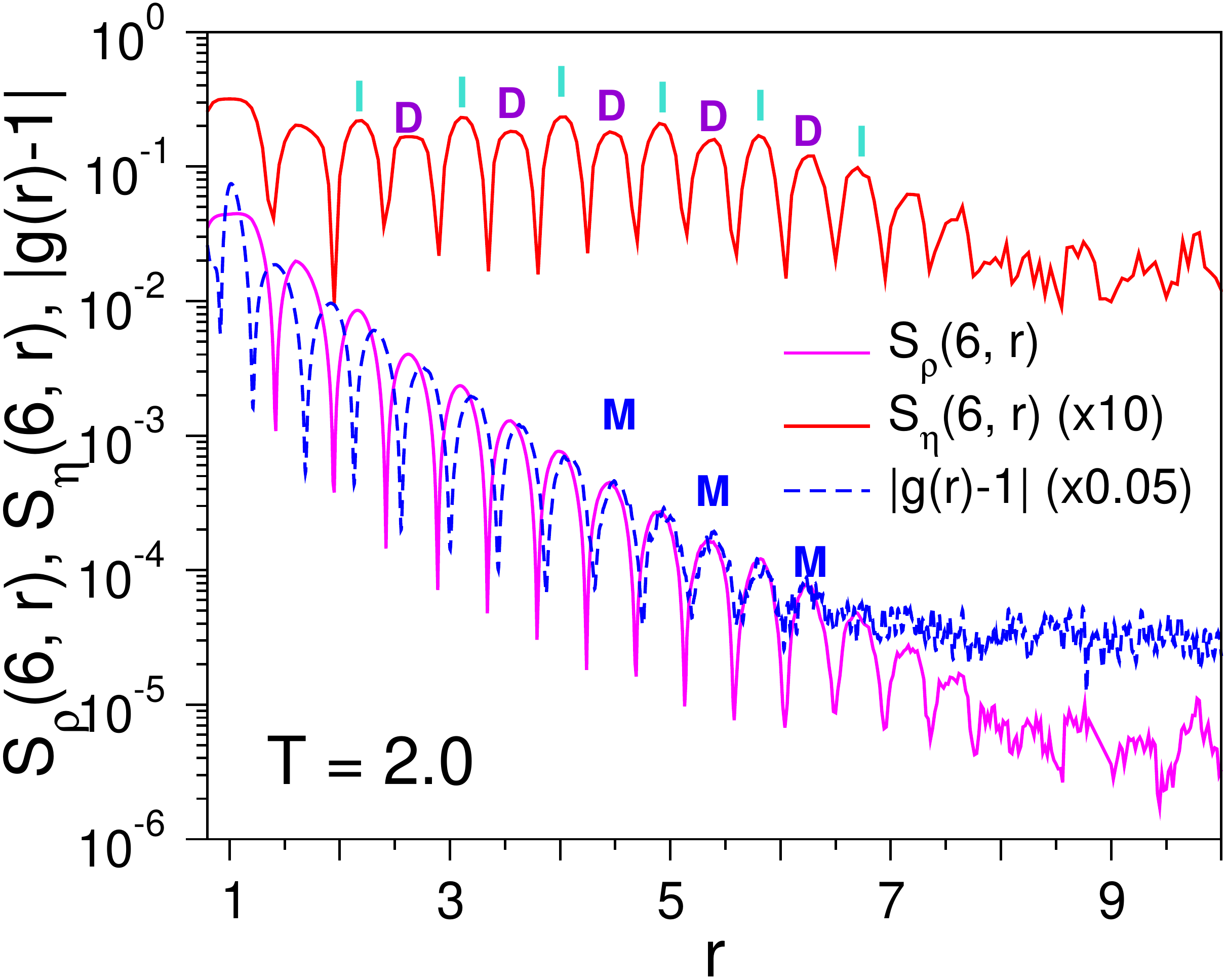}
\caption{{\bf Structural order in the BLJM at $T=2.0$.}
The angular power spectra and radial distribution function
for the BLJM at $T=2.0$.
The power spectrum $S_\rho(6,r)$ (magenta curve)
shows an exponential-like decay as a function of the distance $r$. The
power spectrum for the normalized density distribution, $S_\eta(6,r)$
(red curve), stays large even at intermediate $r$.
For $r \gtrsim 4.0$ the high/low maxima in $S_\eta(r)$, labeled I and D,
coincide with the minima/maxima (labeled M) in $|g(r)-1|$ (blue line).
This up-down behavior is related to the alternating
icosahedral/dodecahedral symmetry in the distribution of the particles
when $r$ is increased.
}
\label{fig2_sff_gr_T2.0}
}
\end{figure}

\noindent
{\bf Temperature dependence of the orientational order}\\ 
The results
in the main text were for a temperature at which the systems were
in a moderately supercooled state (silica) while for the BLJM the
considered temperatures spanned a range in which the system was very
fluid (high $T$) to rather viscous (low $T$). It is important to note
that the orientational order that we have identified is not a result
of the systems being supercooled since it can already be clearly seen
at temperatures at which the system is a normal liquid.  This can be
recognized from Fig.\ref{fig1_densityplot_bljm}{\bf d} in the main text where
we show for the BLJM the 3D density distribution for
various values of $r$ at $T=2.0$, i.e.~at a temperature that is more than
twice the melting temperature, which is around 1.0~\cite{pedersen18}. Even at
this high temperature one can clearly identify at intermediate distance
the presence of shells that have icosahedral and dodecahedral symmetry.
This behavior is quantified in \ref{fig2_sff_gr_T2.0} which shows for $T=2.0$
the $r-$dependence of $S_\rho(6,r)$ and $S_\eta(6,r)$. As it
was the case for the lower temperature $T=0.4$ (Fig.~\ref{fig3_srho_gr}) we
find that at intermediate distances $S_\rho(6,r)$ osciallates perfectly
in phase with $g(r)$ while $S_\eta(6,r)$ stays large at
intermediate distances, i.e., there is a noticable orientational ordering.

Since for the case of silica the temperature we consider is about 50\%
above the melting temperature (2000K), it is evident that for this system
the tetrahedral structure is already present at temperatures that are
well above the melting point.

\begin{figure}[ht]
 \center
\includegraphics[width=0.45\columnwidth]{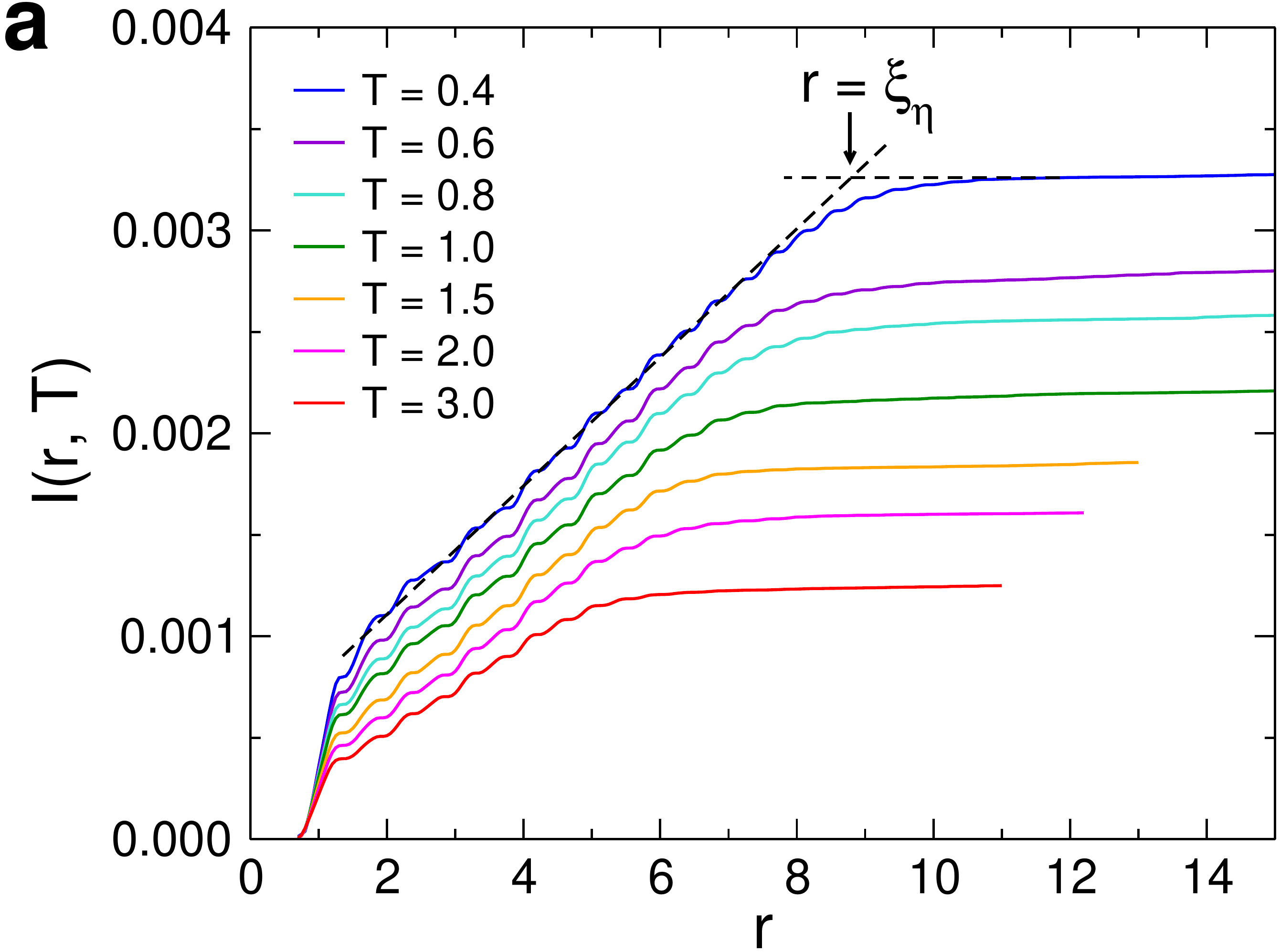}
\includegraphics[width=0.45\columnwidth]{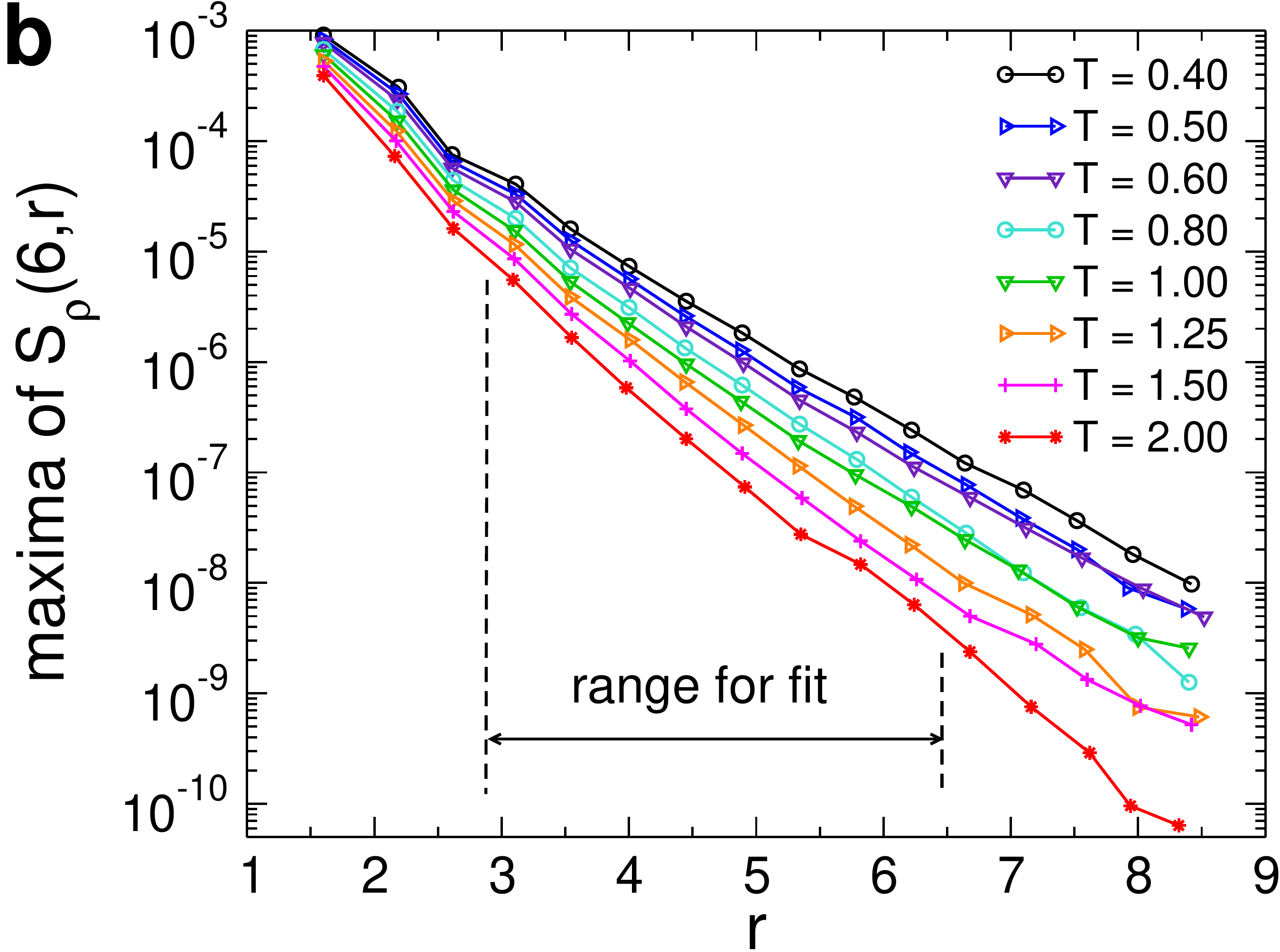}
\includegraphics[width=0.45\columnwidth]{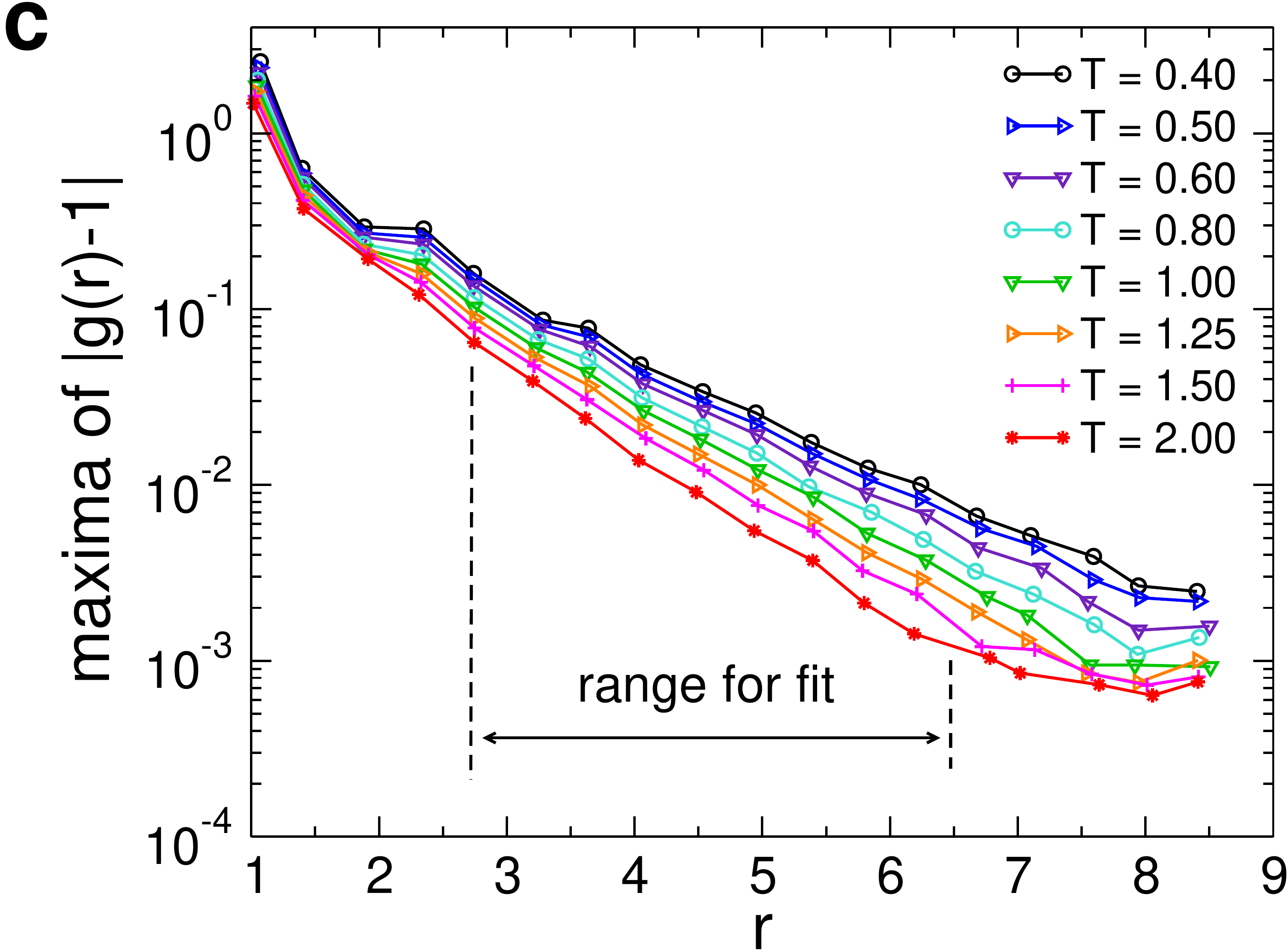}
\includegraphics[width=0.45\columnwidth]{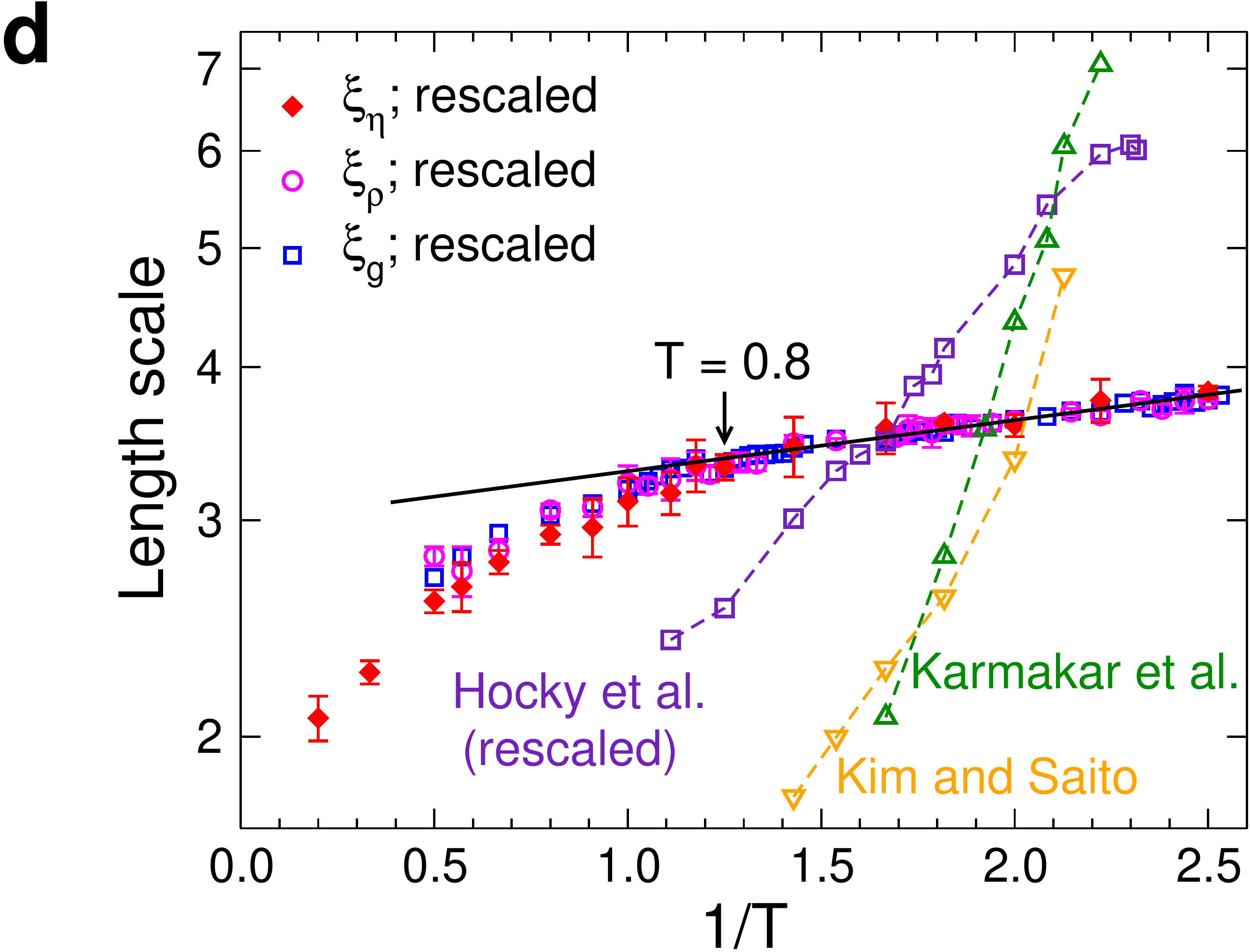}
\caption{{\bf Length scales in the BLJM.}
{\bf a}: $I(r,T)$, the integral of $S_\eta(r)$ for different temperatures.
The length scale $\xi_\eta(T)$ is defined as the crossover point at
which $I(r,T)$ starts to become a constant (see dashed lines).  {\bf b}:
Local maxima of $S_\rho(6,r)$. {\bf c}: Local maxima of $|g(r)-1|$. For
both quantities, the data in the range $2.8 < r < 6.5$ are fitted with
an exponential function to extract the corresponding length scale.
{\bf d}: Different length scales (on log scale) as a function of inverse
temperature: $\xi_\eta$ defined from $S_\eta(r)$ is shown in red and
the inverse of the slope of the exponential decay of $S_\rho(r)$ and
$|g(r)-1)|$ is shown in magenta and blue, respectively.  $\xi_\eta$,
$\xi_\rho$, and $\xi_g$ have been multiplied by a scaling factor of 0.43,
1.29, and 1.29. Error bars are the standard error of the mean of 8 samples.
The line is a guide to the eye to allow to identify the two
temperature dependencies that join at the cross-over temperature around
$T=0.8$. Also included are the dynamical length scale $\xi_4$ as obtained by Karmakar {\it
et al.} and Kim and Saito and a point to set length scale determined by
Hocky {\it et al.} (multiplied by a factor of 3)~$^{35,36,37}$.  }
\label{SI_fig_peak}
\end{figure}

\clearpage

Since the orientational order can be detected at all temperatures,
it is of interest to investigate how this order depends on temperature and in 
the following we present our results for the BLJM.
For this system we have found that at short and intermediate distances
the function $S_\eta(r)$ is basically a constant before it starts
to drop at large distances, see Fig.~\ref{fig3_srho_gr}{\bf a}. 
The distance $\xi_\eta(T)$ at
which $S_\eta(r)$ starts to drop can thus be used to define a static
correlation length. To determine $\xi_\eta$ we have calculated the
integral $I(r,T)=\int_0^r S_\eta(r',T)dr'$ and in \ref{SI_fig_peak}{\bf a}
we plot this quantity as a function of $r$. For small and intermediate
$r$ the integral shows a basically linear increase with $r$, because the
integrand $S_\eta(r)$ is essentially a constant, and once $S_\eta(r)$
starts to decay $I(r,T)$ becomes a constant. Using a fit with two
straight lines this cross-over point can be determined accurately, see
dashed lines in \ref{SI_fig_peak}{\bf a}, giving thus $\xi_\eta(T)$.
Note that the decrease of $S_\eta(r)$ at large distances is due to the
noise in the density field $\rho(\theta,\phi,r)$ and thus the exact
value of $\xi_\eta$ depends on the used statistics, i.e. number of
points used to determine $\rho(\theta,\phi,r)$.  Hence if this noise is
reduced, by increasing the number of particles and points that are used to determine
$\rho(\theta,\phi,r)$, the quasi-constant part of $S_\eta(r)$ at short
and intermediate $r$ will extent to larger distances. As a consequence
the absolute value of $\xi_\eta(T)$ is not a relevant number. However,
if the statistics is kept constant, i.e.~same number of points used 
to calculate $\rho(\theta,\phi,r)$, the $T-$dependence of $\xi_\eta$
is a physically meaningful quantity. The resulting $T-$dependence is plotted in
\ref{SI_fig_peak}{\bf d} and it will be discussed below.

In Figs.~\ref{fig3_srho_gr}{\bf a} and \ref{fig2_sff_gr_T2.0} we have
found that $S_\rho(r)$ has at intermediate and large distances an
exponential dependence on the distance $r$. In \ref{SI_fig_peak}{\bf b}
we show the $r-$dependence of $S_\rho$ for different temperatures. Note
that we plot only the local maxima of the function since these have
been used to fit the data at intermediate and large distances with
an exponential function (see below). From the graph one recognizes
that the slope of the curves decreases with decreasing temperature,
indicating that the associated length scale increases. We obtain this
length scale $\xi_\rho$ by making a fit with an exponential of the form
$S_\rho(r,T) \propto \exp(-r/\xi_\rho(T))$ and include this quantity in
\ref{SI_fig_peak}{\bf d} as well. Also the function $|g(r)-1|$ shows an
exponential decay as a function of $r$ (Fig.~\ref{fig3_srho_gr}{\bf a}
and \ref{fig2_sff_gr_T2.0}) and in \ref{SI_fig_peak}{\bf c} we present
the $r-$dependence of this function for various temperatures. (Again
only the location of the maxima are shown.) Fitting these curves with
an exponential function allows to define a length scale $\xi_g(T)$, the
$T-$dependence of which is included in \ref{SI_fig_peak}{\bf d} as well.\\

\ref{SI_fig_peak}{\bf d} shows the three length scales $\xi_\eta$,
$\xi_\rho$ and $\xi_g(T)$ as a function of inverse temperature and one
recognizes that, after appropriate rescaling, the three length scales
collapse onto each other quite well. In the $T-$range considered, the
scales change by about a factor of 2, i.e.~a relatively modest value.
From the graph one recognizes two regimes: At high $T$ the length scales
increase quickly with decreasing $T$ whereas at low temperatures one finds
a weaker $T-$dependence and which is compatible with $\ln(\xi) \propto
T^{-1}$.  Hence one concludes that a decreasing temperature leads to an
increasing static length scale, in agreement with previous studies that
have documented a weak increase of static length scales in glass-forming
systems, Ref.~\cite{royall_17} and references therein. Surprisingly the
crossover between the two regimes occurs at around $T=0.8$, thus very
close to the so-called ``onset temperature''~\cite{kob_95} at which the
relaxation dynamics of the system crosses over from a normal dynamics
to a glassy one~\cite{binder_11}. This result shows that the change in
the dynamical properties of the system has a counterpart in the statics,
giving hence support to the idea that the latter allows to understand
the former~$^{33}$.

In recent years many efforts have been made to connect the slow
dynamics of glass-forming systems with an increasing static length
scale~\cite{royall_17,royall_15}$^{,34}$. In this context people
have also attempted to identify length scales that are associated
with the {\it dynamics} and to compare these dynamic length scales to the
static ones.  In \ref{SI_fig_peak}{\bf d} we therefore also include
results for the dynamic length scales that have been obtained in previous
works~$^{35,36,37}$ for exactly the same BLJM, namely $\xi_4$
and a point to set length scale. One sees that these dynamic length scales show
a significantly stronger $T-$dependence than the static ones that we
have considered here, a result that is consistent with earlier studies
on this question~\cite{royall_15}.\\[10mm]

==============

\noindent
[29]
T. S. Ingebrigtsen, J. C. Dyre, T. B. Schroder, and C. P. Royall,
Crystallization Instability in Glass-Forming Mixtures.
{\it Phys. Rev. X} {\bf 9}, 031016 (2019).\newline
[30] 
S. Sundararaman, L. Huang, S. Ispas, and W. Kob,
New optimization scheme to obtain interaction potentials for oxide glasses.
{\it J. Chem. Phys.} {\bf 148}, 194504 (2018).\newline
[31] 
S. Plimpton,
Fast parallel algorithms for short-range molecular dynamics.
{\it J. Comp. Phys.} {\bf 117}, 1 (1995).\newline
[33]
W. G\"otze, {\it Complex dynamics of glass-forming liquids:
A mode-coupling theory} (Oxford University Press, Oxford, 2008).\newline
[34] 
A.~Cavagna,
Supercooled liquids for pedestrians.
{\it Phys. Rep.}, {\bf 476}, 51 (2009).\newline
[35] 
S. Karmakara, C. Dasgupta, and S. Sastry,
Growing length and time scales in glass-forming liquids.
{\it Proc. Natl. Acad. Sci. USA} {\bf 106}, 3675 (2009).\newline
[36] 
K.~Kim and S.~Saito,
Multiple length and time scales of dynamic heterogeneities in model
glass-forming liquids: A systematic analysis of multi-point and multi-time
correlations.
{\it J. Chem. Phys.}, {\bf 138}, 12A506 (2013).  \newline
[37] 
G.~M. Hocky, L.~Berthier, W.~Kob, and D.~R.~Reichman,
Crossovers in the dynamics of supercooled liquids probed by an amorphous wall.
{\it Phys. Rev. E}, {\bf 89}, 052311 (2014).  \newline

\clearpage
\noindent{ \bf Movie S1.}\\
This movie shows the density distribution $\rho(\theta,\phi,r)$ as a function
of the distance $r$ (left panel).  The right panel shows the partial
radial distribution function for A-N pairs (blue curve) as well as the
normalized angular power spectrum $S_\eta(6,r)$ (red curve). The center of the vertical
moving bar indicating the radius $r$ shown in the left panel. The
temperature is $T=2.0$.\\[5mm]

\noindent{ \bf Movie S2.}\\
This movie shows the density distribution $\rho(\theta,\phi,r)$ as a function
of the distance $r$ (left panel).  The right panel shows the partial
radial distribution function for A-N pairs (blue curve) as well as the
normalized angular power spectrum $S_\eta(6,r)$ (red curve). The center of the vertical
moving bar indicating the radius $r$ shown in the left panel. The
temperature is $T=0.4$.


\begin{thebibliography}{10}

\bibitem{ashcroft_76}
N. W. Ashcroft and N. D. Mermin, 
{\it Solid State Physics} 
(Holt-Saunders, New York, 1976).

\bibitem{binder_11}
K. Binder and W. Kob,
{\it Glassy Materials and Disordered Solids: An Introduction
to Their Statistical Mechanics}
(World Scientific, Singapore, 2011).

\bibitem{hansen_86}
J. P. Hansen and I. R. McDonald, 
{\it Theory of Simple Liquids}
(Elsevier, Amsterdam, 1986).

\bibitem{salmon_06}
P. S. Salmon, 
Decay of the pair correlations and small-angle scattering for binary liquids and glasses.
{\it J. Phys.: Condens. Matter.} {\bf 18}, 11443 (2006).

\bibitem{tanaka_12}
M. Leocmach and H. Tanaka, 
Roles of icosahedral and crystal-like order in the hard spheres glass transition.
{\it Nat. Comm.} {\bf 3}, 974 (2012).

\bibitem{royall_15}
C. P. Royall and S. R. Williams,
The role of local structure in dynamical arrest.
{\it Phys. Rep.} {\bf 560}, 1 (2015).

\bibitem{coslovich_07}
D. Coslovich and G. Pastore, 
Understanding fragility in supercooled Lennard-Jones mixtures. I. Locally preferred structures.
{\it J. Chem. Phys.} {\bf 127}, 124504 (2007).

\bibitem{dunleavy_15}
A. J. Dunleavy, K. Wiesner, R. Yamamoto, and C. P. Royall,
Mutual information reveals multiple structural relaxation mechanisms in a model glass former.
{\it Nat. Comm.} {\bf 6}, 6089 (2015).

\bibitem{jonsson_88}
H. Jonsson and H. C. Andersen, 
Icosahedral ordering in the Lennard-Jones liquid and glass.
{\it Phys. Rev. Lett.} {\bf 60}, 2295 (1988).

\bibitem{wochner_09}
P. Wochner, C. Gutt, T. Autenrieth, T. Demmer, V. Bugaev, A. D. Ortiz, A. Duri, F. Zontone, 
G. Gr\"ubel, and H. Dosch,
X-ray cross correlation analysis uncovers hidden
local symmetries in disordered matter.
{\it Proc. Natl. Acad. Sci. USA} {\bf 106}, 11511 (2009).

\bibitem{malins_13}
A. Malins, J. Eggers, C. P. Royall, S. R. Williams, and H. Tanaka, 
Identification of long-lived clusters and their link to slow dynamics in a model glass former.
{\it J. Chem. Phys.} {\bf 138}, 12A535 (2013).

\bibitem{ma_11}
Y. Q. Cheng and E. Ma,
Atomic-level structure and structure-property relationship in metallic glasses.
{\it Prog. Mater. Sci.} {\bf 56}, 379 (2011).

\bibitem{miracle_04}
D. B. Miracle, A structural model for metallic glasses.
{\it Nat. Mater.} {\bf 3}, 697 (2004).

\bibitem{xia_17}
C. Xia, J. Li, B. Kou, Y. Cao, Z. Li, X. Xiao, Y. Fu, T. Xiao, L. Hong, J. Zhang, W. Kob, and Y. Wang,
Origin of non-cubic scaling law in disordered granular packing.
{\it Phys. Rev. Lett.} {\bf 118},  238002 (2017).
 
\bibitem{fang_10}
X. W. Fang, C. Z. Wang, Y. X. Yao, Z. J. Ding, and K. M. Ho,
Atomistic cluster alignment method for local order mining in liquids and glasses.
{\it Phys. Rev. B} {\bf 82}, 184204 (2010).

\bibitem{fang_11}
X. W. Fang, C. Z. Wang, S. G. Hao, M. J. Kramer, Y. X. Yao,
M. I. Mendelev, Z. J. Ding, R. E. Napolitano, and K. M. Ho,
Spatially resolved distribution function and the medium-range order in metallic liquid and glass.
{\it Scient. Reps.} {\bf 1}, 194 (2011).

\bibitem{kelton_10}
K. F. Kelton, A. L. Greer,
{\it Nucleation in Condensed Matter: Applications in Materials and Biology}
(Elsevier, Amsterdam, 2010).

\bibitem{royall_17}
C. P. Royall and W. Kob, 
Locally favoured structures and dynamic length scales in a simple glass-former.
{\it J. Stat. Mech.: Theo. Exp.} {\bf 2}, 024001 (2017).

\bibitem{adam_65}
G. Adam and J. H. Gibbs, 
On the temperature dependence of cooperative relaxation properties in glass-forming liquids.
{\it J. Chem. Phys.} {\bf 43}, 139 (1965).

\bibitem{rfot}
X. Y. Xia and P. G. Wolynes, 
Fragilities of liquids predicted from the random first order transition theory of glasses.
{\it Proc. Natl. Acad. Sci. USA}. {\bf 97}, 2990 (2000).

\bibitem{chandler_10}
D.~Chandler and J.~P. Garrahan,
Dynamics on the way to forming glass: Bubbles in space-time.
{\it Annual Review of Physical Chemistry}, 61, 191 (2010).

\bibitem{kob_95}
W.~Kob and H.~C. Andersen, 
Testing mode-coupling theory for a supercooled binary Lennard-Jones mixture I: The van Hove correlation function.
{\it Phys. Rev. E} {\bf 51}, 4626 (1995).

\bibitem{horbach99}
J. Horbach and W. Kob, 
Static and Dynamic Properties of a Viscous Silica Melt.
{\it Phys. Rev. B} {\bf 60}, 3169--3181 (1999).

\bibitem{pedersen18}
U. R. Pedersen, T. B. Schroder, and J. C. Dyre,
Phase Diagram of Kob-Andersen-Type Binary Lennard-Jones Mixtures.
{\it Phys. Rev. Lett.} {\bf 120}, 165501 (2018).

\bibitem{kegel_00}
W. K. Kegel and A. van Blaaderen,
Direct Observation of Dynamical Heterogeneities in Colloidal Hard-Sphere Suspensions.
{\it Science} {\bf 287}, 290 (2000).

\bibitem{weeks_00}
E. R. Weeks, J. C. Crocker, A. C. Levitt, A. Schofield, and D. A. Weitz,
Three-dimensional direct imaging of structural relaxation near the colloidal glass transition.
{\it Science} {\bf 287}, 627 (2000).

\bibitem{sherson_10}
J. F. Sherson, C. Weitenberg, M. Endres, M. Cheneau1, I. Bloch, and S. Kuhr,
Single-atom-resolved fluorescence imaging of an atomic Mott insulator.
{\it Nature} {\bf 467}, 68 (2010).

\bibitem{kou_17}
B. Kou, Y. Cao, J. Li, C. Xia, Z. Li, H. Dong, A. Zhang, J. Zhang, W. Kob, and Y. Wang
Granular materials flow like complex fluids.
{\it Nature} {\bf 551}, 360 (2017).

\end{thebibliography}
\end{document}